 \documentclass[12pt,preprint]{aastex}

\usepackage{amsmath}

\shorttitle{Statistical Analysis of XRFs/XRRs}

\shortauthors{Bi, Mao, Liu, \& Bai}

\begin{document}


\title{Statistical Study of the {\it Swift} X-ray Flash and X-ray Rich Gamma-Ray Bursts}


\author{Xiongwei Bi\altaffilmark{1,2}, Jirong Mao\altaffilmark{3,4,2}, Chuanxi Liu\altaffilmark{3,4,5}, and Jin-Ming Bai\altaffilmark{3,4,2}
}

\altaffiltext{1}{Department of Physics, Honghe University, 661199 Mengzi, China}
\altaffiltext{2}{Key Laboratory for the Structure and Evolution of Celestial Objects, Chinese Academy of Sciences, 650011 Kunming, China}
\altaffiltext{3}{Yunnan Observatories, Chinese Academy of Sciences, 650011 Kunming, Yunnan Province, China}
\altaffiltext{4}{Center for Astronomical Mega-Science, Chinese Academy of Sciences, 20A Datun Road, Chaoyang District, 100012 Beijing, China}
\altaffiltext{5}{University of Chinese Academy of Sciences, 100049 Beijing, China}

\email{jirongmao@mail.ynao.ac.cn}

\begin{abstract}
We build a comprehensive sample to statistically describe the properties of X-ray
flashes (XRFs) and X-ray riches (XRRs) from the latest third \textit{Swift} Burst Alert Telescope (BAT3) catalog of Gamma-ray bursts (GRBs).
We obtain 81 XRFs, 540 XRRs, and 394 classical GRBs (C-GRBs). We statistically explore the different properties of the $\gamma$-ray prompt emission, the X-ray emission, the X-ray lightcurve type, the association with supernovae (SNe), and the host galaxy properties for these sources. We confirm that most XRFs/XRRs are long GRBs with low values of peak energy $E^{obs}_{peak}$ and they are low-luminosity GRBs.
XRFs, XRRs, and C-GRBs follow the same $E_{X,iso}$-$E_{\gamma,iso}$-$E_{peak,z}$ correlations. Compared to the classical GRBs, XRFs are favorable to have the association with SN explosions. We do not find any significant differences of redshift distribution and host galaxy properties among XRFs, XRRs, and C-GRBs.
We also discuss some observational biases and selection effects that may affect on our statistical results.
The GRB detectors with wide energy range and low energy threshold are expected for the XRF/XRR research in the future.
\end{abstract}




\section{Introduction} \label{sec:intro}

In addition to the classical long/short dichotomy, gamma-ray bursts (GRBs) have two special subclasses: X-ray flashes (XRFs) and X-ray riches (XRRs). XRFs are the GRBs characterized by the faint signals in the gamma-ray energy band. XRRs, which belong to an intermediate class between XRFs and classical GRBs (C-GRBs), have stronger X-ray emission compared to their gamma-ray emission (e.g, Barraud et al. 2003; Kippen et al. 2003; Amati et al. 2004; Sakamoto et al. 2005, 2008; D'Alessio et al. 2006). The physical origin of XRFs and XRRs are still under debate.

XRF 050406 was proposed as a GRB with prelonged central engine activity (Romano et al. 2006). This long-term activity was also observed in XRF 011030 (Galli \& Piro 2006). GRB jet structure affects on the observed GRB energy release, and the off-axis effect may induce the observed XRFs/XRRs (e.g., Yamazaki et al. 2002; Barraud et al. 2005; Granot et al. 2005; Lamb et al. 2005; Xu et al. 2005; Donaghy 2006; Salafia et al. 2016). Observations have provided some further evidence for the off-axis jet injection (e.g., Bulter et al. 2005; Schady et al. 2006; de Ugarte Postigo 2007; Guidorzi et al. 2009). A dynamic transition with a different GRB jet opening angle may also be important to link C-GRBs and XRFs (Mizuta et al. 2006).
Alternatively, thermal emission has been thought to be a possible component in the strong X-ray emission of GRBs. Ramirez-Ruiz (2005) proposed a photospheric model that can be used to interpret the dominated X-ray emission of XRFs. Pe'er et al. (2006) calculated the details of the photospheric component of the XRF prompt emission spectrum.
XRFs can also be the indicators of the orphan GRB afterglows (Urata et al. 2015). These clues naturally lead to one suggestion that XRFs are low-luminosity
GRBs (e.g., Virgili et al. 2009).
Moreover, it has been found that supernova (SN) explosions can be associated with XRFs (e.g., XRF 020903, Bersier et al. 2006).

XRF 050215B was the first XRF observed by {\it Swift} (Levan et al. 2006), and {\it Swift} observational statistics can be well applied to study the physical origins of XRFs and XRRs (Gendre et al. 2007). Sakamoto et al. (2008) built one dataset provided by the \textit{Swift} Burst Alert telescope (BAT) observation from 2004 December to 2006 September. From that sample, they obtained 10 XRFs and 97 XRRs among a total of 158 GRBs.
They studied the prompt emission properties and the X-ray afterglow emission characteristics for XRFs, XRRs, and C-GRBs.
Some distinct differences between XRFs and C-GRBs in the prompt emission and the X-ray afterglow emission have been illustrated.
This exploration encourages us to comprehensively investigate the physical properties of XRFs/XRRs, the relation between XRFs/XRRs and C-GRBs, and the GRB central engine from the statistical point of view. Thus, a large GRB sample is necessary.

We utilize the latest third \textit{Swift}-BAT3 catalog (Lien et al. 2016), which contains 1104
GRBs detected from 2004 December 17 to 2016 December 2, to systematically investigate the statistical properties of XRFs and XRRs.
For each GRB, the catalog provides trigger time, coordinates, redshift, GRB duration time $T_{90}$, spectral models for spectral fitting, spectral photon index, observed peak energy $E_{peak}$, and fluence in difference energy bands.
From this catalog, we classify the possible XRFs/XRRs and build a sample to comprehensively analyze the differences of XRFs/XRRs and C-GRBs. We identify
81 XRFs, 540 XRRs, and 394 C-GRBs included in the sample and statistically analyze their observational characteristics. We examine the possible associations between XRFs/XRRs and SNe. The host galaxy properties of XRFs are also presented.
Some observational biases and selection effects are mentioned.

This paper is organized as follows. We classify XRFs, XRRs, and C-GRBs in Section 2. In Section 3, we present the different
properties of XRFs, XRRs, and C-GRBs, respectively. A discussion is provided in Section 4, and conclusions are listed in Section 5.
We adopt the cosmological parameters as $H_{0}=70~\rm{km~s^{-1}~Mpc}^{-1}$, $\Omega_{\Lambda}=0.7$, and $\Omega_{M}=0.3$.
The quoted errors are at the 90\% confidence level unless stated otherwise.

\section{Sample Selection} \label{sec:sample}
We classify XRFs, XRRs, and C-GRBs using the criteria provided by Sakamoto et al. (2008). The definitions of XRF, XRR, and C-GRB based
on the fluence ratio of $S(25-50~\rm{keV})$ and $S(50-100~\rm{keV})$ are:
\begin{equation}
\begin{split}
S(25-50~\rm{keV})/S(50-100~\rm{keV})\leq 0.72~~~~(\rm{C}-\rm{GRB}),\\
0.72<S(25-50~\rm{keV})/S(50-100~\rm{keV})\leq 1.32~~~~(\rm{XRR}),\\
S(25-50~\rm{keV})/S(50-100~\rm{keV})>1.32~~~~(\rm{XRF}).\\
\end{split}
\end{equation}
We obtain 81 XRFs, 540 XRRs, and 394 C-GRBs in the {\it Swift}-BAT3 catalog by the selection condition of Equation (1). We summarize the selection results in Table 1.
We further consider three additional conditions. First, in order to compare the
prompt emission and the X-ray afterglow emission of each XRF/XRR, we exclude 89 sources that have no {\it Swift} X-ray Telescope (XRT) data. Thus, we do not clarify these sources
as XRFs, XRRs, or C-GRBs. We also specify that 13 sources in our sample have no GRB duration $T_{90}$
numbers. Thus, we cannot clarify them as long-duration GRBs (L-GRBs, defined by $T_{90} \geq 2$ s) or short-duration GRBs (S-GRBs, defined
by $T_{90} \leq 2$ s).
Second, we also select GRBs that have the photon index $\alpha_{PL} < -2.0$ with a power-low fitting
in the BAT3 dataset and identify them as XRFs. Third, we note the sources having values of $E^{obs}_{peak}$, which can be found in the BAT3 dataset. We also check whether these selected sources have $E^{obs}_{peak}$ values in other datasets. Finally, we list XRFs and XRRs in Table 2.

The distributions with the fluence ratio of $S$(25-50 keV)/$S$(50-100 keV) for the total 1015 GRBs in the BAT3 catalog are shown in Figure 1.
XRFs, XRRs, and C-GRBs in the sample have the fractions of $(8.0\pm 0.9)$\%, $(53.2\pm 2.2)$\%, and $(38.8\pm 1.9)$\%, respectively.
The smaller sample given by Sakamoto et al. (2008) contains 158 GRBs. There are 10 XRFs, 97 XRRs, and 51 C-GRBs.
XRFs, XRRs, and C-GRBs in their sample have the fractions of $(6.3\pm 2.0)$\%, $(61.4\pm 6.2)$\%, and $(32.3\pm 4.5)$\%, respectively.
It seems that our classified results are roughly consistent with those of Sakamoto et al. (2008).
Here, we pay attention to four special cases:
(1) GRB 050219B was identified as XRR by Sakamoto et al. (2008), while it is classified as C-GRB in our work. (2) GRB 050815 was identified as XRR by Sakamoto et al. (2008), while it is classified as XRF in our work.
(3) There are 10 GRBs (GRB 050824, GRB 060512, GRB 060923B, GRB 060926, GRB070714A, GRB070721A, GRB080218B, GRB080515, GRB080520, and GRB081007) that have no fluences of $S$(25-50 keV) and/or $S$(50-100 keV) in BAT3 catalog, and we find the fluences of $S$(25-50 keV) and/or $S$(50-100 keV) from the {\it Swift}-BAT2 catalog (Sakamoto et al. 2011). (4) There are 11 XRFs (XRF 050406, XRF 050416A, XRF 050819, XRF 060428B, XRF 060805A, XRF 061218, XRF 070126, XRF 080218B, XRF 080315, XRF 080822B, and XRF 160525A) show a fluence ratio of $S$(25-50 keV)/$S$(50-100 keV) larger than 3.0\footnote{XRF 080315 has the largest fluence ratio of $53.8\pm 143.0$ but with large error. This source has no X-ray afterglow from {\it Swift}-XRT detection. Due to the lack of X-ray nondetection and the marginal BAT detection (Page \& Gehrels 2008), we include this source in Table 1, 2 and 3, but we exclude it in all Figures of this paper. We also do not find any other notable issues for this burst.}.

We plot the fluence ratio $S$(25-50 keV)/$S$(50-100 keV) versus the BAT-observed GRB duration $T_{90}$ in Figure 2. We calculate the fractions of L-GRBs and S-GRBs for XRFs, XRRs, and C-GRBs in our sample, respectively. Our findings are as follows: (1) For XRFs, there are 70 L-GRBs, 3 S-GRBs, and 8 duration-unclear sources, and the fractions are $(86.4\pm 10.3)$\%, $(3.7\pm 2.1)$\%, and $(9.9\pm 3.5)$\%, respectively. (2) For XRRs, there are 509 L-GRBs, 27 S-GRBs, and 4 duration-unclear sources, and the fractions are $(94.3\pm 4.2)$\%, $(5.0\pm 1.0)$\%, and $(0.7\pm 0.4)$\%, respectively. (3) For C-GRBs, there are 328 L-GRBs, 65 S-GRBs, and 1 duration-unclear source, and the fractions are $(83.2\pm 4.6)$\%, $(16.5\pm 2.0)$\%, and $(0.3\pm 0.3)$\%, respectively. We note that XRFs and XRRs have less S-GRB proportion compared with C-GRBs\footnote{There are only three XRFs (XRF 090417A, XRF 110112A, and XRF 140622A) that are S-GRBs. GRB 110112A has no host galaxy evidence (Fong et al. 2013; Tunnicliffe et al. 2014). We do not find any other notable information for the short-duration XRFs.}.
In the meantime, we also report the fraction of XRFs, XRRs, and C-GRBs for L-GRB and S-GRB classes. Three S-GRBs as XRFs have the fraction of $(3.2\pm 1.8)$\%. Twenty-seven S-GRBs as XRRs have the fraction of $(28.4\pm 5.5)$\%. Sixty-five S-GRBs as C-GRBs have the fraction of $(68.4\pm 8.5)$\%.
Seventy L-GRBs as XRFs have the fraction of $(7.7\pm 0.9)$\%. Five hundred and nine L-GRBs as XRRs have the fraction of $(56.1\pm 2.5)$\%. Three hundred and twenty-eight L-GRBs
as C-GRBs have the fraction of $(36.2\pm 2.0)$\%.

\section{Statistical Analysis} \label{sec:analysis}

\subsection{The Prompt Emission Properties} \label{subsec:prompt}
We collect $E^{obs}_{peak}$ values of GRBs from the literature (e.g., Amati et al. 2008, 2009; Sakamoto et al. 2008, 2011; Grupe et al. 2013; D'Avanzo et al. 2014; Liang et al. 2015; Lien et al. 2016; Zaninoni et al. 2016), and we obtain the $E^{obs}_{peak}$  values for 77 XRFs, 460 XRRs, and 265 C-GRBs.
The fluence ratio $S$(25-50 keV)/$S$(50-100 keV) versus $E^{obs}_{peak}$ is shown in Figure 3.
We clearly see the different occupied regions of XRFs, XRRs, and C-GRBs in the Figure.
It was shown in the fluence ratio-$E^{obs}_{peak}$ plot provided by Sakamoto et al. (2008) a gap of $S$(25-50 keV)/$S$(50-100 keV) fluence ratio from 0.8 to 1.2,
and Sakamoto et al. (2008) suggested that this gap is the result of selection effects.
However, we do not find this gap in Figure 3, because we take a large sample from the BAT3 catalog.
We further show the different $E^{obs}_{peak}$ distributions for XRFs, XRRs, and C-GRBs in Figure 4.
In order to quantitatively distinguish the different $E^{obs}_{peak}$ properties to XRFs, XRRs, and C-GRBs,
we use a nonparametric two-sample Kolmogorov-Smirnov (K-S) test to examine the different $E^{obs}_{peak}$ distributions for the XRF/XRR samples, the XRR/C-GRB samples, and the XRF/C-GRB samples, respectively.
Because the K-S probability numbers are very small (the $P$-values are far less than 0.0001), we confirm that the $E^{obs}_{peak}$ distributions among XRFs, XRRs, and C-GRBs have significant differences.

The $E^{obs}_{peak}$ distribution of XRFs ranges from 0.9 keV to 80.0 keV,
with a mean value of 24.3$\pm$1.6 keV.
The $E^{obs}_{peak}$ distribution of XRRs ranges from 1.2 keV to 1780.0 keV,
with a mean value of 105.6$\pm$5.3 keV.
The $E^{obs}_{peak}$ distribution of C-GRBs ranges from 64.6 keV to 2602.8 keV,
with a mean value of 257.7$\pm$14.1 keV.
It is clear that XRFs and XRRs have smaller $E^{obs}_{peak}$ values compared with C-GRBs.
Our results are consistent with those of Sakamoto et al. (2005, 2008).
We confirm that XRFs and XRRs release their prompt energies mostly in the X-ray band.

We also investigate the correlation between peak energy $E^{obs}_{peak}$ and the fluence $S$(15-150 keV) for all GRBs in our sample.
The data are plotted in Figure 5.
In principle, the effect of the data errors should be taken into account when we perform the correlation fitting. In this paper, we adopt the maximum likelihood method that has been well applied for the $E_{\rm{peak,z}}$-$E_{\rm{\gamma,iso}}$ correlation fitting given by Amati et al. (2008). We use the maximum likelihood method and obtain the correlation fitting as
$\log(E^{obs}_{peak})$(keV)=$(2.96\pm 0.13)+(0.16\pm 0.02)\log[S(15-150)]\rm{keV}$ with the extrinsic scatter $\sigma=0.39\pm 0.01$.
We also plot the correlation of $\log(E^{obs}_{peak})$(keV)=$(5.46\pm 0.25)+(0.62\pm0.14)\log[S(15-150$keV)] that was given by Sakamoto et al. (2008). We see that our fitting result is different from that of Sakamoto et al. (2008).
In order to clarify the difference of this correlation among XRFs, XRRs, and C-GRBs, we separate C-GRBs as one group and put XRRs and XRFs as the other group. We obtain the fitting for C-GRBs as $\log(E^{obs}_{peak})$(keV)=$(3.01\pm 0.13)+(0.12\pm 0.02)\log[S(15-150)]\rm{keV}$ with the extrinsic scatter $\sigma=0.26\pm 0.01$, and the fitting for XRRS and C-GRBs as $\log(E^{obs}_{peak})$(keV)=$(2.67\pm 0.16)+(0.13\pm 0.03)\log[S(15-150)]\rm{keV}$ with the extrinsic scatter $\sigma=0.36\pm 0.01$. Therefore, although the difference between XRFs and XRRs/C-GRBs is clear, it seems no significant difference between XRFs/XRRs and C-GRBs because XRRs and C-GRBs has large overlap region seen in Fig. 5.

\subsection{The Observed Properties with Redshift}\label{subsec:observed properties}
The redshift distributions of the XRFs, XRRs, and C-GRBs are shown in Figure 6.
Using the K-S test to the redshift distributions for the XRFs and XRRs samples, the XRRs and C-GRBs samples, and the XRFs and C-GRBs samples, we find that K-S test probabilities are $P=0.13$, $P=0.36$, and $P=0.13$, respectively. The K-S test results confirm that there are not significant differences among XRFs, XRRs, and C-GRBs for the redshift distribution.

We also plot the BAT-observed duration $T_{90}$ and the fluence $S$(15-150 keV) as a function of redshift in Figure 7 and Figure 8, respectively.
We do not find significant differences of $T_{90}$ and $S$(15-150 keV) distributions among XRFs, XRRs, and C-GRBs, and we do not see significant redshift evolutions of $T_{90}$ and $S$(15-150 keV) for XRFs, XRRs, and C-GRBs.

\subsection{The Correlations among $E_{X,iso}$, $E_{\gamma,iso}$, and $E_{peak,z}$} \label{subsec:isotropic prompt}
There is a universal correlation among the isotropic prompt energy $E_{\gamma,iso}$ emitted in the rest frame $1-10^{4}$ keV energy band, the rest frame
energy peak of the prompt emission energy spectrum $E_{peak,z}$ in which $E_{peak,z}$=$(1+z)E^{obs}_{peak}$, and
the X-ray energy emitted in the rest frame $0.3-10$ keV energy band $E_{X,iso}$ for GRBs (e.g., Bernardini et al. 2012; Margutti et al. 2013; D'Avanzo et al. 2014; Zaninoni et al. 2016).
In order to check whether XRFs, XRRs, and C-GRBs in our sample follow this correlation, respectively,
we collect the $E_{\gamma,iso}$ and $E_{X,iso}$ data of the GRBs in our sample from Amati et al. (2008), Margutti et al. (2013), and Liang et al. (2015).
First, the relation between $E_{X,iso}$ and $E_{\gamma,iso}$ is shown in Figure 9. We perform the maximum likelihood method and obtain the fitting of
$(E_{X,iso})=(13.31\pm 2.73)+(0.73\pm 0.05)log(E_{\gamma,iso})$ with the extrinsic scatter of $\sigma=0.57\pm 0.04$.
The $E_{X,iso}$-$E_{\gamma,iso}$ relation is consistent with that derived by Margutti et al. (2013).
Second, we also investigate the correlation between $E_{peak,z}$ and $E_{\gamma,iso}$. The result is shown in Figure 10.
The correlation fitted by the maximum likelihood method is $\log(E_{peak,z})$(keV)=$(2.17\pm 0.04)+(0.46\pm 0.03)\log[E_{\gamma,iso}/(10^{52}erg)]$ with the extrinsic scatter of $\sigma=0.26\pm 0.02$.
Our result is consistent with the correlation reported by Amati (2006).
Third, in Figure 11, we present the relation between $E_{X,iso}$ and $E_{peak,z}$ by the maximum likelihood method with the fitting of
$log(E_{X,iso})=(49.69\pm 0.27)+(0.75\pm 0.11)log(E_{peak,z})$, and the extrinsic scatter is $\sigma=0.73\pm 0.04$.
Our result is consistent with that of Margutti et al. (2013).
Finally, the $E_{X,iso}$-$E_{\gamma,iso}$-$E_{peak,z}$ relation is shown in Figure 12.
The result with the fitting of
$log(E_{X,iso})=(4.78\pm 2.79)+(0.92\pm 0.06)(logE_{\gamma,iso}-0.60log(E_{peak})$ and the extrinsic scatter of $\sigma=0.44\pm 0.04$
produced by the maximum likelihood method
is consistent with that of Margutti et al. (2013) as well.

We also examine the difference between XRFs/XRRs and C-GRBs from the above correlations. We separate XRFs/XRRs and C-GRBs as two groups. From Figure 9, we obtain $(E_{X,iso})=(9.85\pm 2.77)+(0.79\pm 0.05)log(E_{\gamma,iso})$ with the extrinsic scatter of $\sigma=0.68\pm 0.07$ for C-GRBs and $(E_{X,iso})=(13.45\pm 2.53)+(0.73\pm 0.05)log(E_{\gamma,iso})$ with the extrinsic scatter of $\sigma=0.39\pm 0.04$ for XRRs/XRFs. Thus, we do not find significant difference in this correlation.
we analyze the data in Figure 10 that the relation of $\log(E_{peak,z})$(keV)=$(2.83\pm 0.06)+(0.11\pm 0.05)\log[E_{\gamma,iso}/(10^{52}erg)]$ with the extrinsic scatter of $\sigma=0.31\pm 0.04$ is for C-GRBs and the relation of $\log(E_{peak,z})$(keV)=$(2.11\pm 0.03)+(0.41\pm 0.04)\log[E_{\gamma,iso}/(10^{52}erg)]$ with the extrinsic scatter of $\sigma=0.23\pm 0.03$ is for XRRs/XRFs. Thus, we clearly see the difference between XRRs/XRFs and C-GRBs in this correlation.
From Figure 11, We obtain the relation of $log(E_{X,iso})=(48.07\pm 1.03)+(1.24\pm 0.35)log(E_{peak,z})$ with the extrinsic scatter of $\sigma=0.86\pm 0.09$ for C-GRBs and the relation of $log(E_{X,iso})=(49.46\pm 0.28)+(0.90\pm 0.12)log(E_{peak,z})$ with the extrinsic scatter of $\sigma=0.62\pm 0.05$ for XRRs/XRFs. Thus, it seems that XRRs/XRFs and C-GRBs have no significant difference in this correlation.
Finally, from Figure 12, we obtain the relation of $log(E_{X,iso})=(5.67\pm 3.36)+(0.90\pm0.07)(logE_{\gamma,iso}-0.6log(E_{peak})$ with the extrinsic scatter of $\sigma=0.54\pm 0.08$ for C-GRBs and the relation of $log(E_{X,iso})=(4.10\pm 2.88)+(0.93\pm 0.06)(logE_{\gamma,iso}-0.60log(E_{peak})$ with the extrinsic scatter of $\sigma=0.36\pm 0.05$. Therefore, we do not find any significant difference between XRRs/XRFs and C-GRBs in this correlation.

In order to investigate the $\gamma$-ray isotropic-equivalent luminosity ($L_{\gamma,iso}$) distributions for XRFs, XRRs, and C-GRBs,
we collect the $L_{\gamma,iso}$ values of the GRBs in our sample from D'Avanzo et al. (2014), Liang et al. (2015), and Cano et al. (2017).
We obtained 55 sources with $L_{\gamma,iso}$ values, including 5 XRFs, 26 XRRs, and 24 C-GRBs. The $L_{\gamma,iso}$ distributions are shown in Figure 13.
Our results are as follows: the $L_{\gamma,iso}$ values have the range from $2.60\times10^{46}$ to $6.89\times10^{50}~\rm{erg~s^{-1}}$ for XRFs;
the $L_{\gamma,iso}$ values are from $1.03\times10^{49}$ to $3.51\times10^{52}~\rm{erg~s^{-1}}$ for XRRs;
and the $L_{\gamma,iso}$ values are from $1.20\times10^{50}$ to $1.78\times10^{53}~\rm{erg~s^{-1}}$ for C-GRBs.
We find that XRFs and XRRs have lower $L_{\gamma,iso}$ values than C-GRBs. This indicates that XRF sources are low-luminosity GRBs.

\subsection{The X-Ray Lightcurve Shapes}\label{subsec:x-ray afterglow}
In order to investigate the X-ray afterglow properties of XRFs, XRRs, and C-GRBs, we simply examine the types of X-ray afterglow light curves for XRFs, XRRs, and C-GRBs
from {\it Swift}-XRT GRB lightcurve repository\footnote{http://www.swift.ac.uk/xrt$\_$curves/}.
The type definitions of X-ray afterglow light curve given by Margutti et al. (2013) are as follows: type 0 (simple power law), type \uppercase\expandafter{\romannumeral1} (broken power law), type \uppercase\expandafter{\romannumeral2} (broken power law plus power-law decay), and Type \uppercase\expandafter{\romannumeral3} (double broken power laws).
We further check the X-ray lightcurve types of XRFs, XRRs, and C-GRBs in our sample.
Our results are as follows: (1) There are 33 XRFs having the XRT light curves. For these XRFs, the proportions of Type 0, Type \uppercase\expandafter{\romannumeral1},
Type \uppercase\expandafter{\romannumeral2}, and Type \uppercase\expandafter{\romannumeral3} are $(14.7\pm 6.6)$\%, $(44.1\pm 11.4)$\%, $(32.4\pm 9.8)$\%, and $(8.8\pm 5.1)$\%, respectively.
(2) There are 198 XRRs having the XRT light curves. For these XRRs, the proportions of Type 0, Type \uppercase\expandafter{\romannumeral1},
Type \uppercase\expandafter{\romannumeral2}, and Type \uppercase\expandafter{\romannumeral3} are $(17.4\pm 2.9)$\%, $(30.8\pm 3.9)$\%, $(45.3\pm 4.7)$\%, and $(6.5\pm 1.8)$\%, respectively.
(3) There are 135 C-GRBs having the XRT light curves. For these C-GRBs, the proportions of Type 0, Type \uppercase\expandafter{\romannumeral1},
Type \uppercase\expandafter{\romannumeral2}, and Type \uppercase\expandafter{\romannumeral3} are $(25.2\pm 4.3)$\%, $(44.4\pm 5.7)$\%, ($27.4\pm 4.5)$\%, and $(3.0\pm 1.5)$\%, respectively.
The statistical results are summarized in Table 3.
According to these results, we find that there are not significant differences of the XRT lightcurve type among XRFs, XRRs, and C-GRBs.

\subsection{Investigation of the Association of XRF/XRR with Supernova}\label{subsec:supernova}
If we propose that XRFs and XRRs are low-luminosity GRBs, it is reasonable to consider the possible association between XRFs/XRRs and SNe (Soderberg et al. 2005; Woosley \& Bloom 2006)\footnote{Although we focus on low-luminosity GRBs and
XRFs in this paper, we note that high-luminosity GRBs may also have SN association. For example, bright GRB 130427A is
associated with SN 2013cq (Maselli et al. 2014; Melandri et al. 2014; Vestrand et al. 2014).}. An example is XRF 060218 that is associated with SN 2006aj (Pian et al. 2006). We take the statistical results from Hjorth \& Bloom (2012), and Cano (2013), and Cano et al. (2017).
Twenty-three GRBs in our sample are associated with the SN explosion. These GRBs include 6 XRFs (XRF 050416A, XRF 050824, XRF 060218/SN 2006aj, XRF 070419A, XRF 081007/SN 2008hw, and XRF 100316D/SN 2010bh), 14 XRRs (XRR 050525A/SN 2005nc, XRR 060729, XRR 060904B, XRR 090618, XRR 091127/SN 2009nz, XRR 101219B/SN 2010ma, XRR 101225A, XRR 111228A, XRR 120422A/SN 2012bz, XRR 120714B/SN 2012eb, XRR 120729A, XRR 130215A/SN 2103ez, XRR 130831A/SN 2013fu, and XRR 150818A), and 3 C-GRBs (GRB 080319B, GRB 111209A/SN 2011kl, and GRB 130427A/SN 2013cq). Our statistical results are shown in Table 4.
It seems that XRFs and XRRs are more favorable to link with SN events than C-GRBs.

\subsection{Host Galaxy Properties}\label{subsec:host galaxy}
The host galaxies of XRFs were investigated in the work of Bloom et al. (2003). Here, we investigate the host galaxy properties for the XRFs, XRRs, and C-GRBs in the BAT3 catalog.
We pay attention to several parameters of GRB host galaxies from the GRB Host Studies (GHostS) database\footnote{see the webpage http://www.grbhosts.org/. We note that GRB 120422 is classified as XRF in GhostS, but it is identified as XRR in this paper.}.
The physical quantities of GRB host galaxy are stellar mass ($M^{\ast}$), metallicity ($Z$), and star formation rate (SFR).
The distributions of $M^{\ast}$ for XRFs, XRRs, and C-GRBs are shown in Figure 14. This Figure includes 6 XRFs, 17 XRRs, and 28 C-GRBs.
We cannot find the significant differences among XRFs, XRRs, and C-GRBs, as we estimate the K-S test probabilities between
XRFs and XRRs ($P=0.67$), XRRs and C-GRBs ($P=0.30$), and XRFs and C-GRBs ($P=0.28$).
The distributions of metallicity $Z$ for XRFs, XRRs, and C-GRBs are shown in Figure 15. We obtain metallicity values of 5 XRFs, 6 XRRs, and 13 C-GRBs.
It is hard to distinguish the differences among XRFs, XRRs, and C-GRBs. The K-S test probabilities between XRFs and XRRs, between XRRs and C-GRBs, between XRFs and C-GRBs are $P=0.97$, $P=0.44$, and $P=0.90$, respectively.
The distributions of SFR are shown in Figure 16. 4 XRFs, 11 XRRs, and 22 C-GRBs are included. The K-S test probabilities are $P=0.27$ (between XRFs and XRRs),
$P=0.09$ (between XRRs and C-GRBs), and $P=0.03$ (between XRFs and C-GRBs). Therefore, it seems that there is an SFR difference between XRFs and C-GRBs. Here, we also note that only four XRFs have SFR values. This limitation prevents us for the further investigation.

GRB host galaxies are usually considered to be low-mass, low-metallicity, and star-forming galaxies (Christensen et al. 2004; Fynbo et al. 2009; Savaglio et al. 2009). However, we see that some GRBs are hosted in massive and/or high-metallicity galaxies (e.g., Hashimoto et al. 2015). Mao (2010) proposed a possible redshift evolution of GRB host galaxies from the theoretical point of view. From the observational point of view, one survey of {\it Swift}-GRB host galaxy has recently been performed (Perley et al. 2016).
We hope that more GRB host properties of XRFs, XRRs, and C-GRBs can be explored for the further statistical analysis in the future.

\subsection{Observational Biases and Selection Effects}\label{subsec:selection}
We should mention some selection effects and observational biases that may affect on our statistical results.
First, the GRB prompt emission spectrum is usually fitted by the Band function \citep{band93}.
However, the detection energy range of {\it Swift}-BAT is $15-350$ keV, such that the spectral
fitting is performed in the narrow energy range. The $E^{obs}_{peak}$ determination is from the cutoff power-law spectral model.
Thus, the $E^{obs}_{peak}$ values in this sample might be different from those obtained from other space telescope detections with a wide energy range.
Because GRB detections in a large energy range with a low energy threshold are required to accurately measure the $E_{peak}^{obs}$ numbers, some future sensitive telescopes, such as the Space Variable Objects Monitor (SVOM) and the Einstein Probe (EP), are expected (Yuan et al. 2017).
Second, GRB redshift values are determined by the spectral observations in the optical band. Thus, we cannot ignore the fact that many XRFs/XRRs have no redshift determinations. Hence, the XRF/XRR quantities related to the redshift cannot be determined. One incomplete sample may have bias on the redshift distribution (e.g., Fiore et al. 2007).
Third, the observations for GRB host galaxies are also complicated. The detection of the high-redshift GRB host galaxies is one challenge. For example, Basa et al. (2012) performed the host galaxy search for three GRBs with $z>5$ using the {\it Hubble Space Telescope}, and they did not find any evidence of high-redshift GRB hosts.
Although Mao et al. (2010) presented the possible redshift evolution of GRB host properties, the GRB redshift distribution with the cosmic star formation
has some biased effects (e.g., Dainotti et al. 2015).

\section{Summary}
We present a comprehensively statistical analysis to study the XRF/XRR properties in the {\it Swift}-BAT3 catalog.
We have obtained 81 XRFs and 540 XRRs in our sample.
We have analyzed the properties of $\gamma$-ray prompt emission, X-ray emission, X-ray light curve type, association with SNe, and host galaxy properties for XRFs, XRRs, and C-GRBs. We list the major findings as follows: (1) Most XRFs/XRRs have low values of $E^{obs}_{peak}$. We confirm that XRFs/XRRs mainly release their energy in the X-ray band, and they are low-luminosity GRBs. (2) Most XRFs/XRRs are long-duration GRBs. (3) XRFs, XRRs, and C-GRBs follow the same $E_{X,iso}$-$E_{\gamma,iso}$-$E_{peak,z}$ correlations. (4) We do not find any differences of redshift distributions among XRFs, XRRs, and C-GRBs in our sample.
(5) XRFs seem to favor the association with SN explosions. (6) We find marginal but interesting evidence that
different SFRs are shown between XRRs/XRFs and C-GRBs.

Although we see some differences between XRFs/XRRs and C-GRBs in some correlation studies and statistic results, we notice that
the properties of XRFs, XRRs, and C-GRBs do not show a sharp difference. We confirm that XRFs and XRRs are belong to GRBs.
However, the physical origin of XRF/XRR is still unclear. The jet off-axis effect is traditionally applied to explain the observational phenomena of XRFs/XRRs. The constraints of the jet beaming and the opening angle were already proposed (Rhoads 1999; Frail et al. 2001). However, it is not the case that each GRB with the jet off-axis is XRF/XRR. For example, GRB 080710 with the observational evidence of the jet off-axis (Kr\"{u}hler et al. 2009) is classified as C-GRB in this paper.
On the other hand, the direct measurements to identify the jet off-axis evidence cannot be performed for each GRB. The jet beaming angle statistics related to the study of XRF/XRR is expected (Gao \& Dai 2010).
It is suggested that the GRB thermal component in the X-ray band can be one possible reason to explain XRF/XRR energy release. As an example, GRB 090618, identified as XRR in this paper, had a detection of thermal X-ray emission by {\it Swift} X-ray telescope, and this XRR is associated with SN explosion (Cano et al. 2011; Page et al. 2011). Starling et al. (2012) presented 11 {\it Swift}-detected GRBs with optical SN explosions, and the thermal X-ray signatures were clearly identified. However, compared to the XRFs/XRRs listed in this paper, the observed GRBs with the thermal emission that have optical SN explosion evidence are still very rare.
From some recent theoretical modeling analysis, thermal emission may regulate the GRB spectral peak energy (Beloborodov 2013). Photospheric models can reproduce the GRB thermal emission in the $\gamma$-ray band (Vurm et al. 2013).
We expect further observational cases of the GRB thermal emissions, although most of
XRR and XRF spectra are still nonthermal.
Finally, the GRB detectors with wide energy range and low energy threshold are expected especially for the study of XRFs and XRRs in the future.

\acknowledgments
The work is financially supported by the Open Project Program of the Key Laboratory for the Structure and Evolution of Celestial Objects,
Chinese Academy of Sciences (grant No. OP201501), the Scientific Research Foundation of the Education Department of Yunnan Province (grant No. 2014Y463), and
the National Natural Science Foundation of China (11404103). J. M. is supported by the National Natural Science Foundation of China
(11673062 and 11661161010), the Hundred Talent Program, the Major Program of the Chinese
Academy of Sciences (KJZD-EW-M06), and the Oversea Talent Program of Yunnan Province. J.-M. B. is partly supported by the National Natural Science Foundation of China (11433004) and the Ministry of Science and Technology of China (2016YFA0400700).

\clearpage
\begin{figure}
\plotone{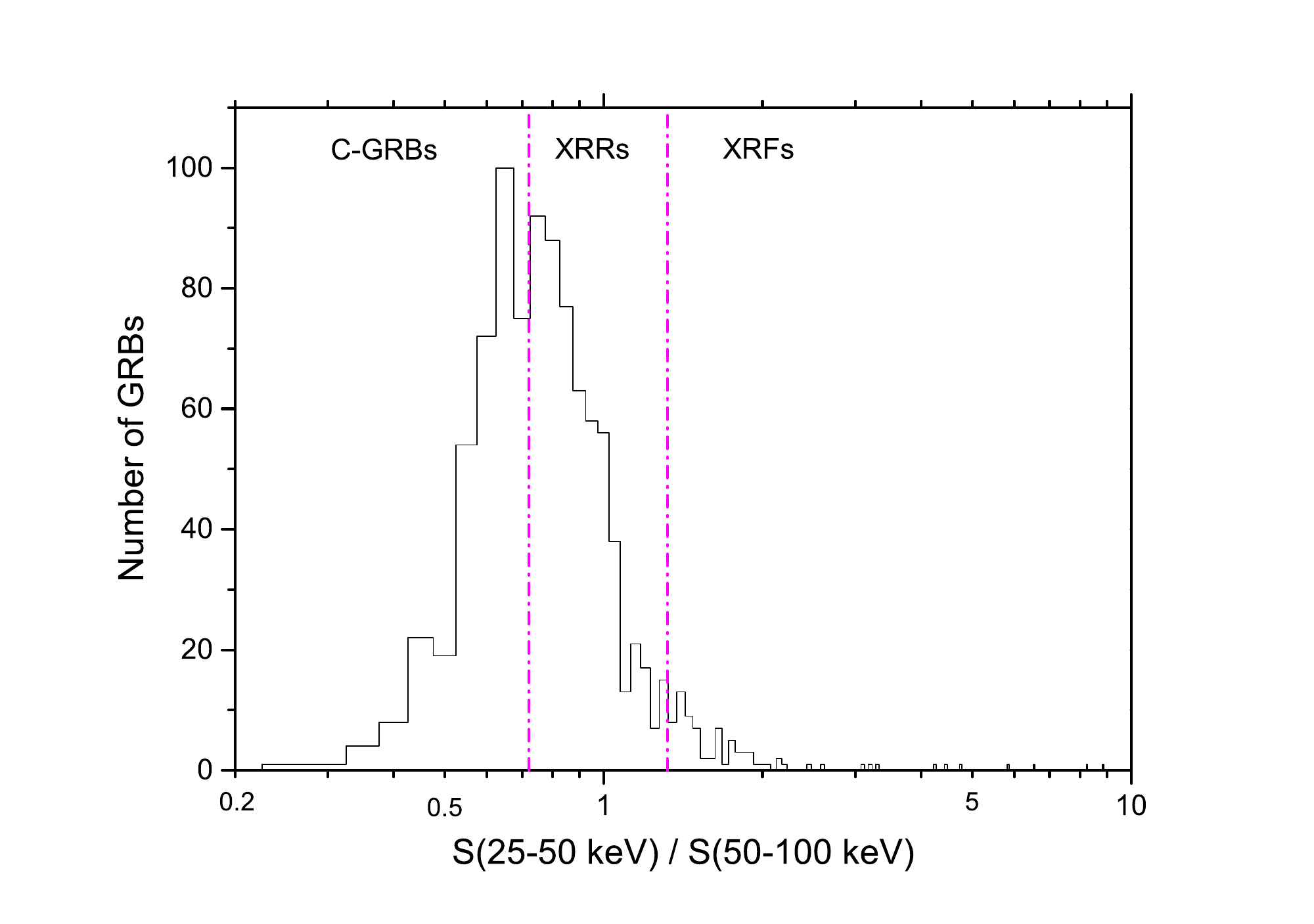}
\caption{Distributions of the fluence ratio $S$(25-50 keV)/$S$(50-100 keV) of GRBs in {\it Swift}-BAT3 sample. The dashed lines show the distribution borders between C-GRBs and XRRs, and between XRRs and XRFs, respectively. Here, we list the sources with the fluence ratio larger than 3.0: XRF 050819 (the ratio is $3.08\pm 0.99$), XRF 050406 (the ratio is $3.15\pm 1.22$), XRF 060805A (the ratio is $3.29\pm 2.20$), XRF 080218B (the ratio is $4.24\pm 3.08$), XRF 060428B (the ratio is $4.44\pm 1.80$), XRF 080822B (the ratio is $4.74\pm 2.86$), XRF 050416A (the ratio is $5.83\pm$2.39), XRF 160525A (the ratio is $6.50\pm 5.47$), XRF 061218 (the ratio is $8.25\pm 10.66$), and XRF 070126 (the ratio is $8.82\pm 9.46$).
\label{fig1}}
\end{figure}

\begin{figure}
\plotone{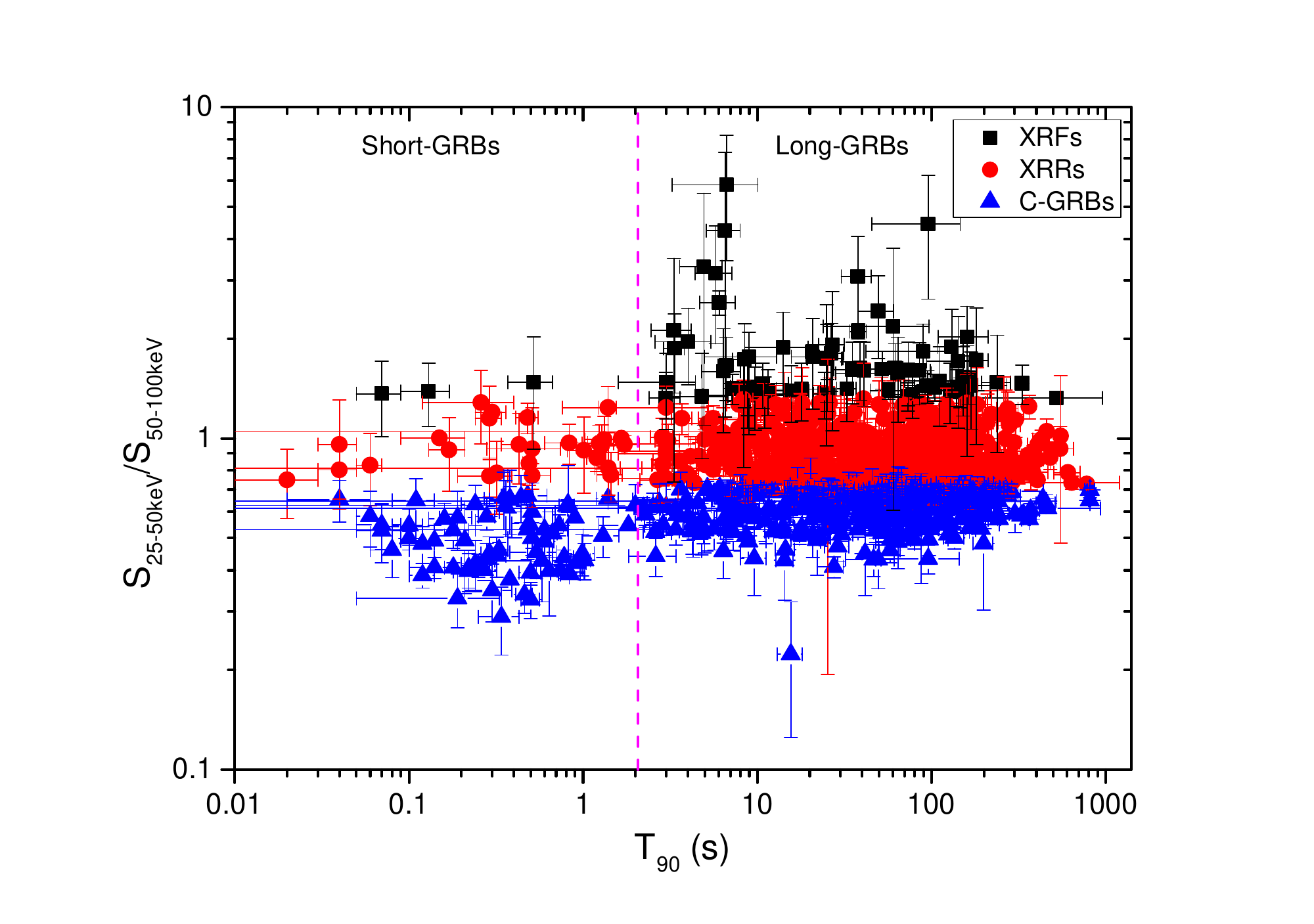}
\caption{The fluence ratio of $S$(25-50 keV)/$S$(50-100 keV) vs. the duration $T_{90}$ of GRBs in the {\it Swift}-BAT3 sample. XRFs, XRRs, and C-GRBs are marked as black squares, red dots, and blue triangles, respectively. The dashed line shows the distribution border between long GRBs and short GRBs. \label{fig2}}
\end{figure}

\begin{figure}
\plotone{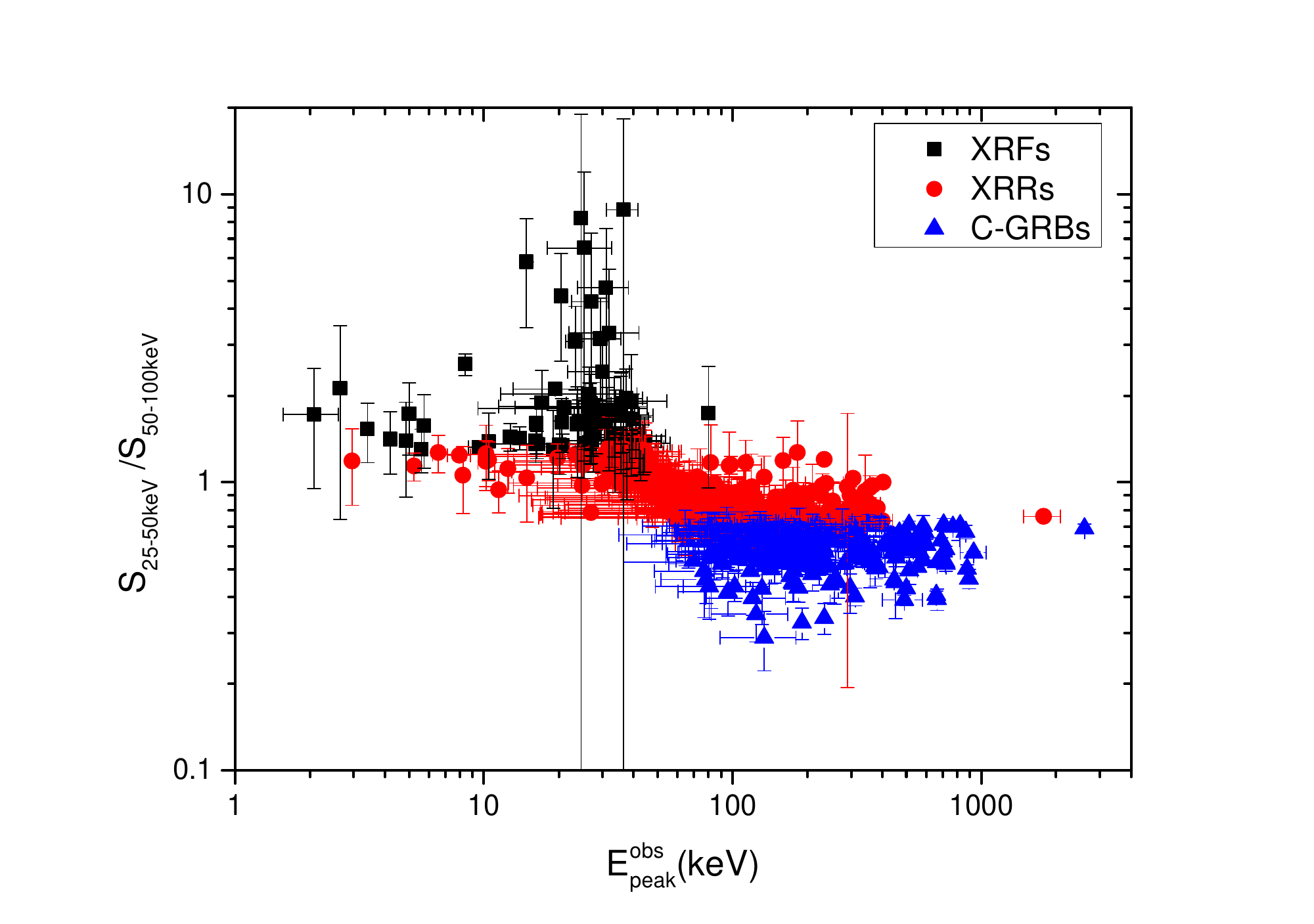}
\caption{Fluence ratio of $S$(25-50 keV)/$S$(50-100 keV) versus the peak energy $E^{obs}_{peak}$ for XRFs, XRRs, and C-GRBs. XRFs, XRRs, and C-GRBs are marked as black squares, red dots, and blue triangles, respectively. XRF 061218 and XRF 070126 have large fluence ratio error bars, and they are not included in Figure 2 because they have no $T_{90}$ values. \label{fig4}}
\end{figure}

\begin{figure}
\plotone{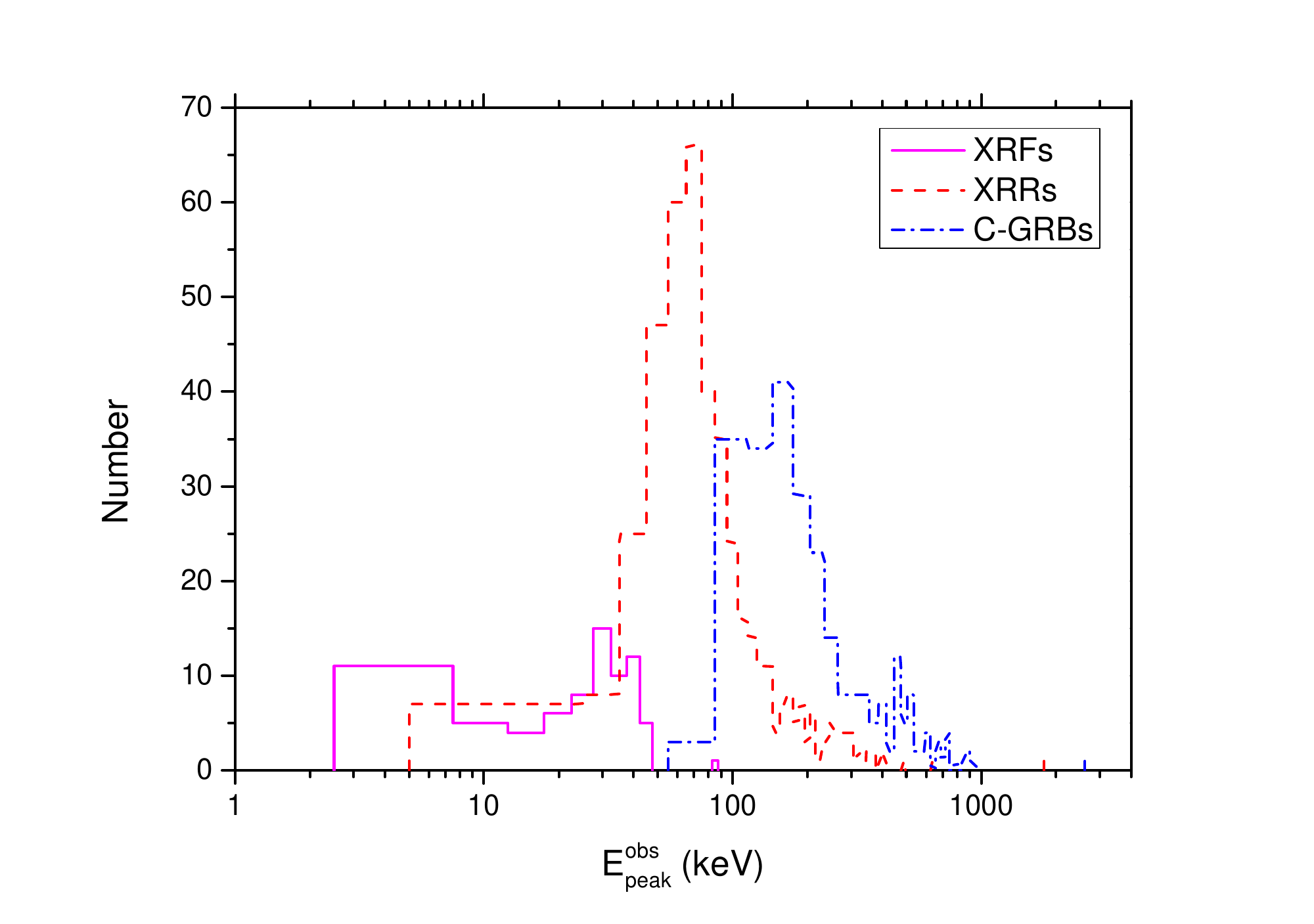}
\caption{$E^{obs}_{peak}$ distribution for XRFs, XRRs, and C-GRBs. XRFs, XRRs, and C-GRBs are marked as the pink solid line, the red dashed line, and the blue dash-dotted line, respectively. \label{fig4}}
\end{figure}

\begin{figure}
\plotone{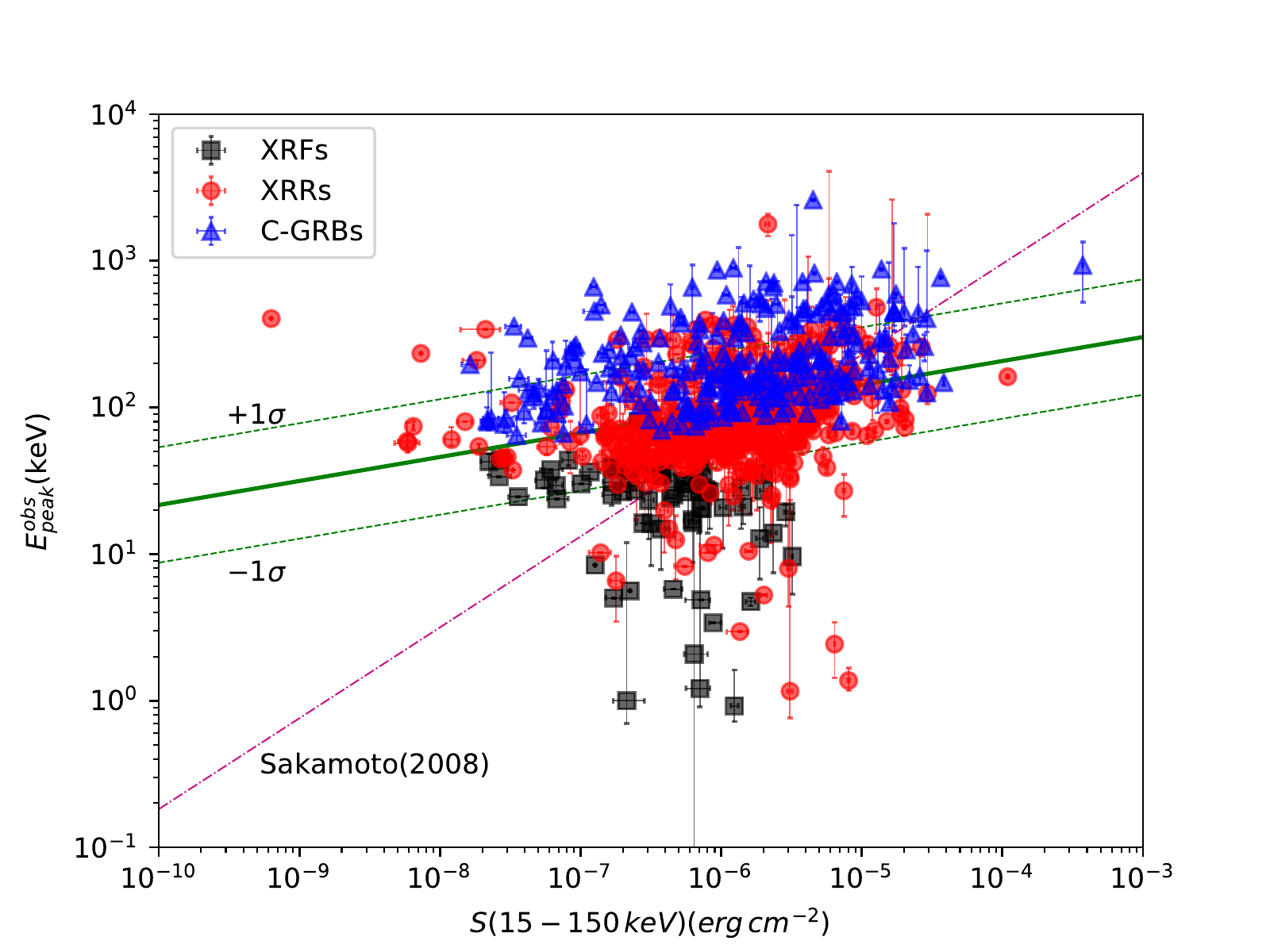}
\caption{Relationship of the 15-150 keV fluence and the $E^{obs}_{peak}$ for XRFs, XRRs, and C-GRBs. The green solid line is the best fit to the data with the function of
$\log(E^{obs}_{peak})$(keV)=$(2.96\pm 0.13)+(0.16\pm 0.02)\log[S(15-150)]\rm{keV}$,
and the green dashed lines are marked for the 1 $\sigma$ regions. The extrinsic scatter $\sigma=0.39\pm 0.01$. The pink dashed-dotted line is the best fit to the data without taking into account the errors reported by Sakamoto et al. (2008), and the function is $\log(E^{obs}_{peak})$(keV)=$(5.46\pm 0.25)+(0.62\pm0.14)\log[S(15-150)$keV]. The XRFs, XRRs, and C-GRBs are marked as black squares, red dots, and blue triangles, respectively. \label{fig5}}
\end{figure}

\begin{figure}
\plotone{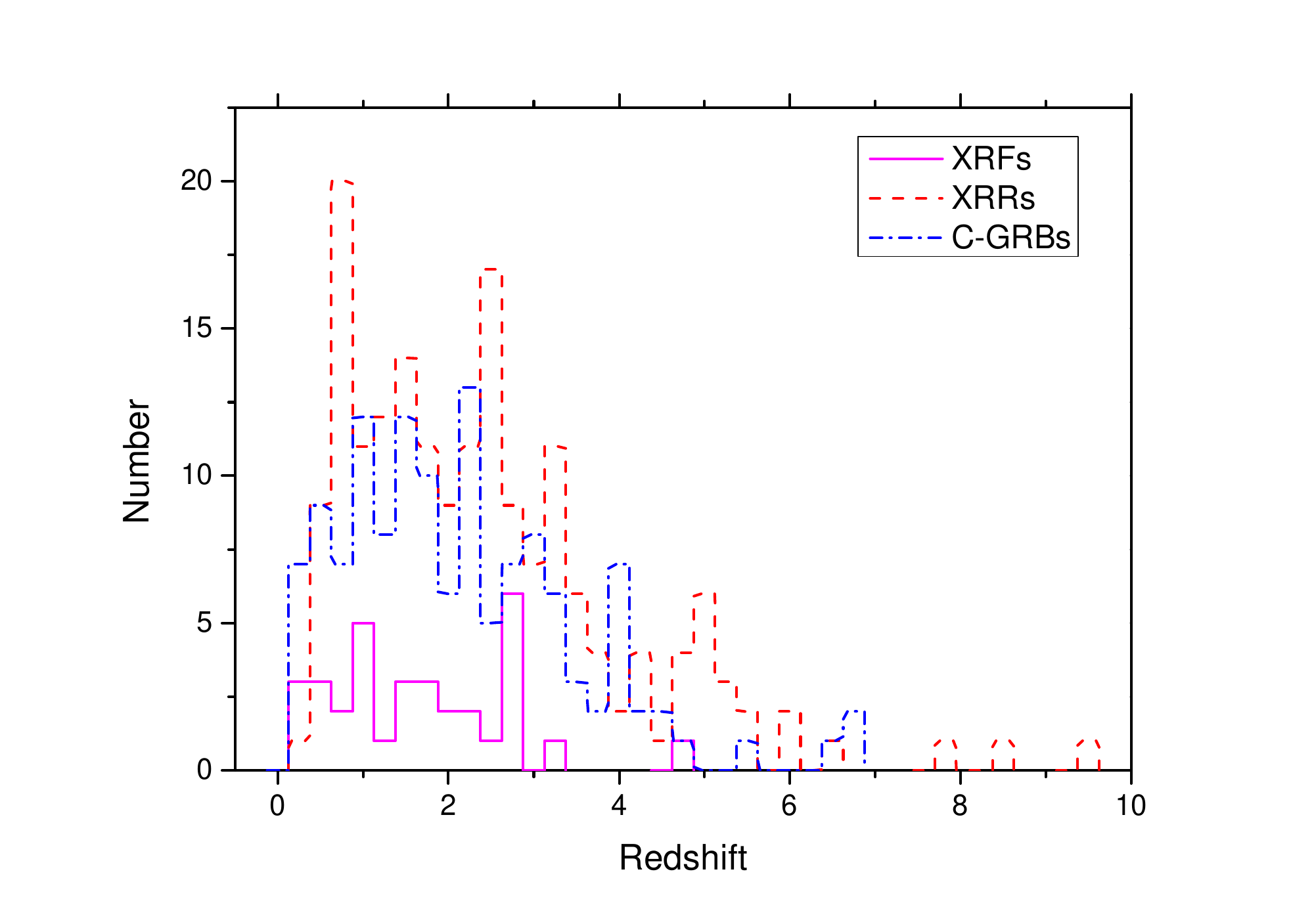}
\caption{Redshift distribution of GRBs in the {\it Swift}-BAT3 sample. XRFs, XRRs, and C-GRBs are marked as the pink solid line, the red dashed line, and the blue dashed-dotted line, respectively. \label{fig6}}
\end{figure}

\begin{figure}
\plotone{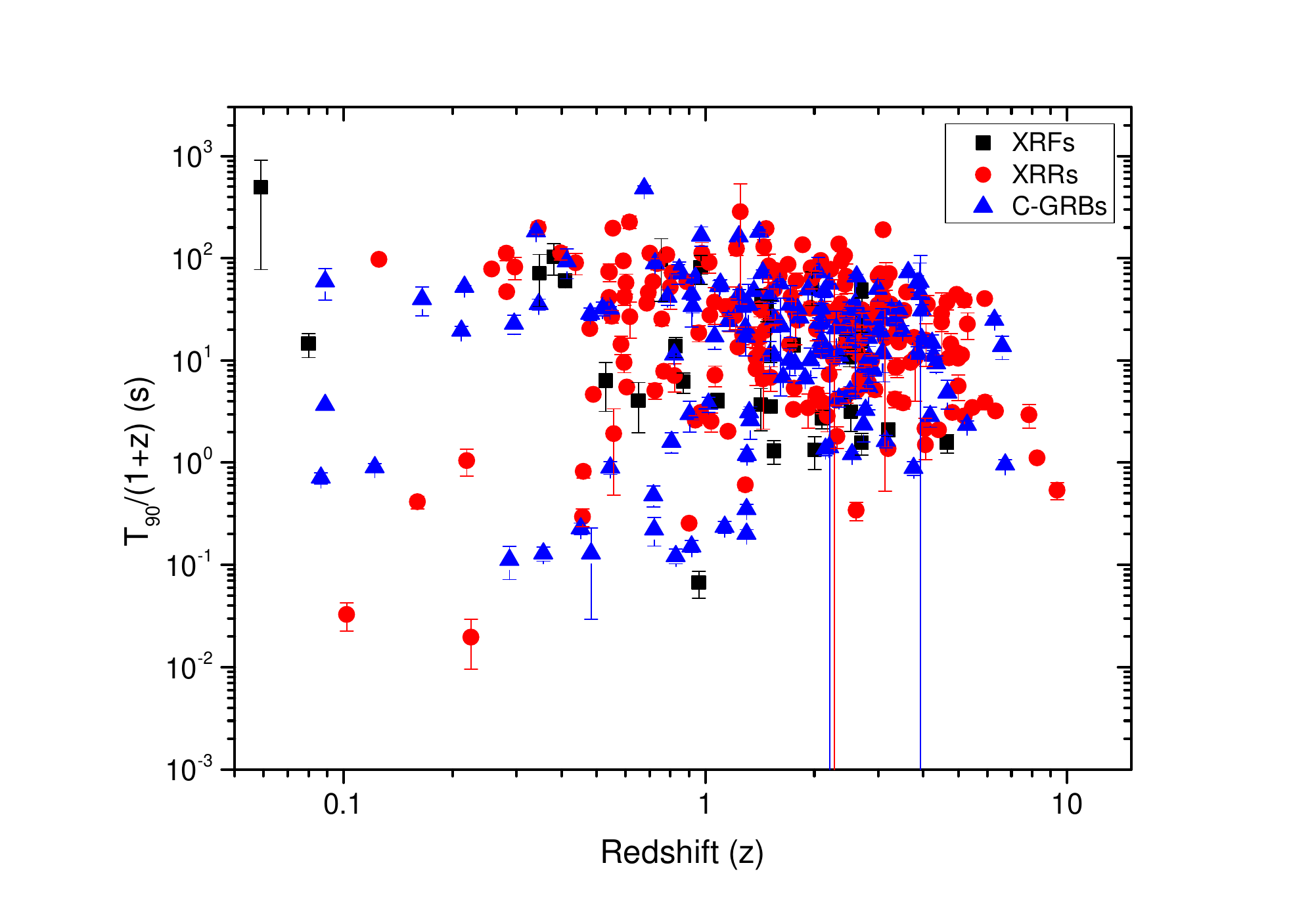}
\caption{GRB duration $T_{90}$ vs. redshift. XRFs, XRRs, and C-GRBs are marked as black squares, red dots, and blue triangles, respectively. \label{fig7}}
\end{figure}

\begin{figure}
\plotone{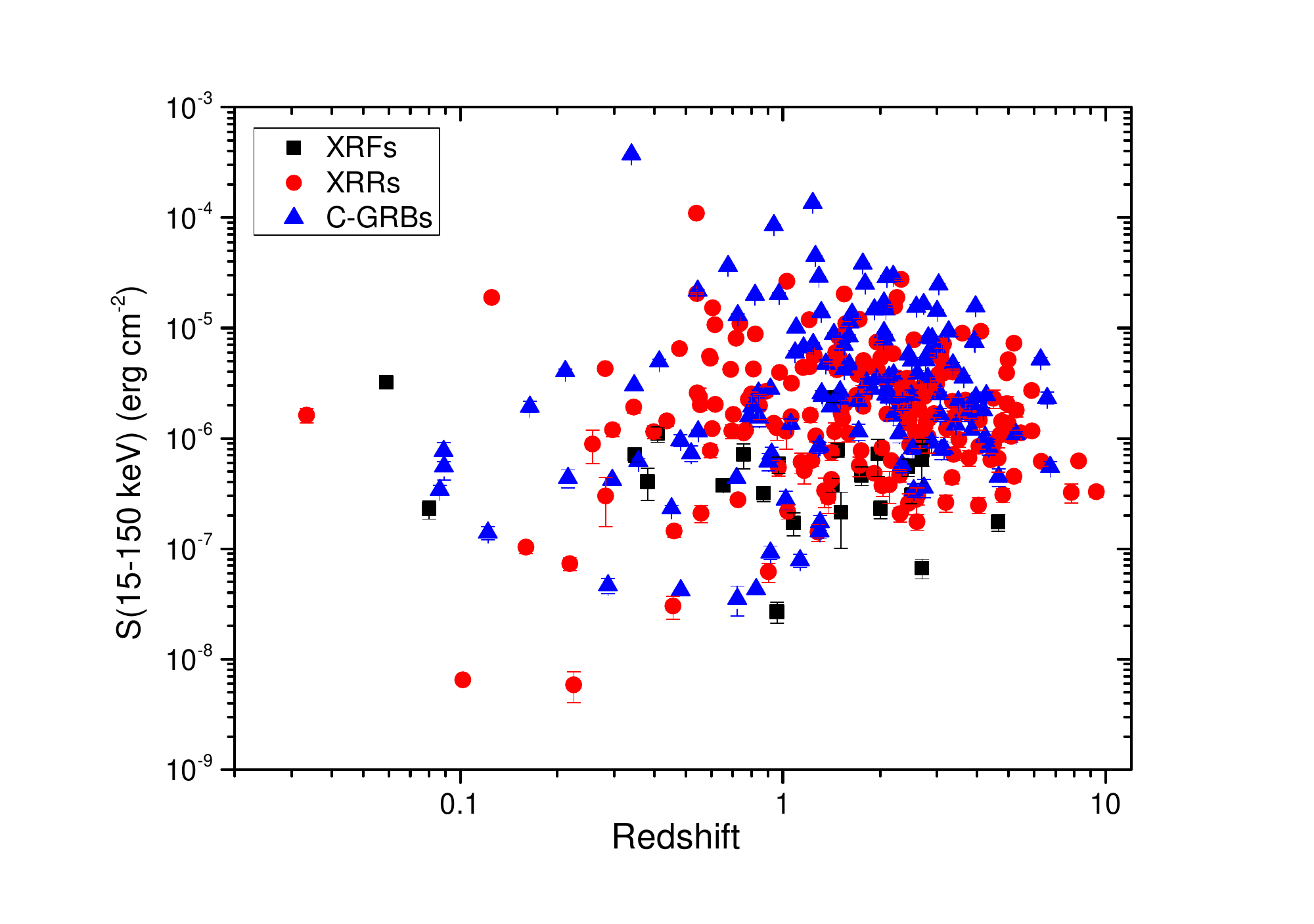}
\caption{GRB fluence in the 15-150 keV band vs. redshift. XRFs, XRRs, and C-GRBs are marked as  black squares, red dots, and blue triangles, respectively. \label{fig8}}
\end{figure}

\begin{figure}
\plotone{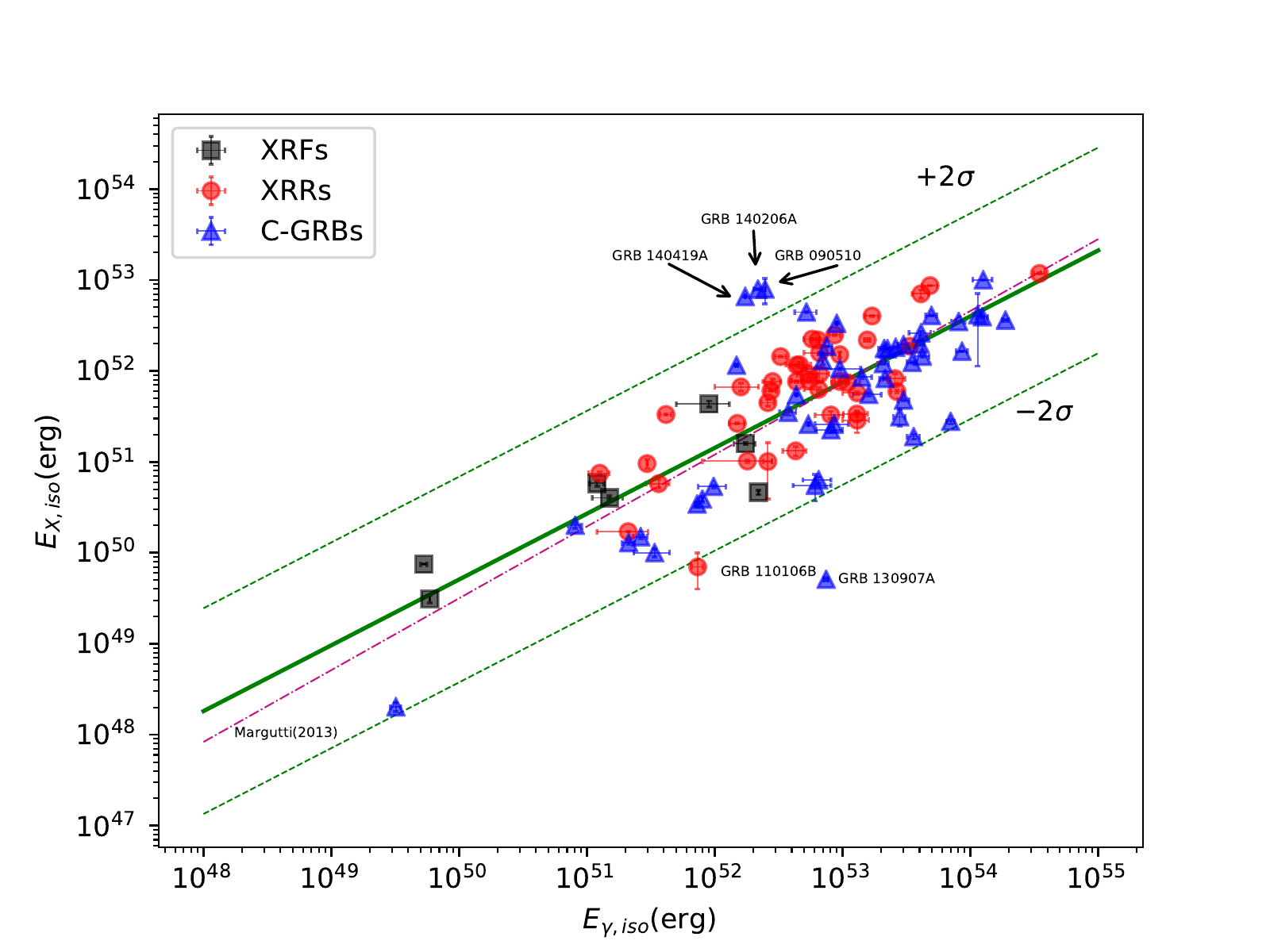}
\caption{Correlation between $E_{X,iso}$ and $E_{\gamma,iso}$ in the {\it Swift}-BAT3 sample. The green solid line is the best fit with the function of
$(E_{X,iso})=(13.31\pm 2.73)+(0.73\pm 0.05)log(E_{\gamma,iso})$, and the extrinsic scatter is $\sigma=0.57\pm 0.04$.
The green dashed lines are marked for the 2$\sigma$ regions.
The pink dashed-dotted line is the best-fitting function of $log(E_{X,iso})=(10.0\pm 20.6)+(0.79\pm 0.01)log(E_{\gamma,iso})$
reported by Margutti et al.(2013). XRFs, XRRs, and C-GRBs are marked as black squares, red dots, and blue triangles, respectively. The GRBs labeled as outliers in the figure are L-GRBs, and the exception is short GRB 090510. \label{fig9}}
\end{figure}

\begin{figure}
\plotone{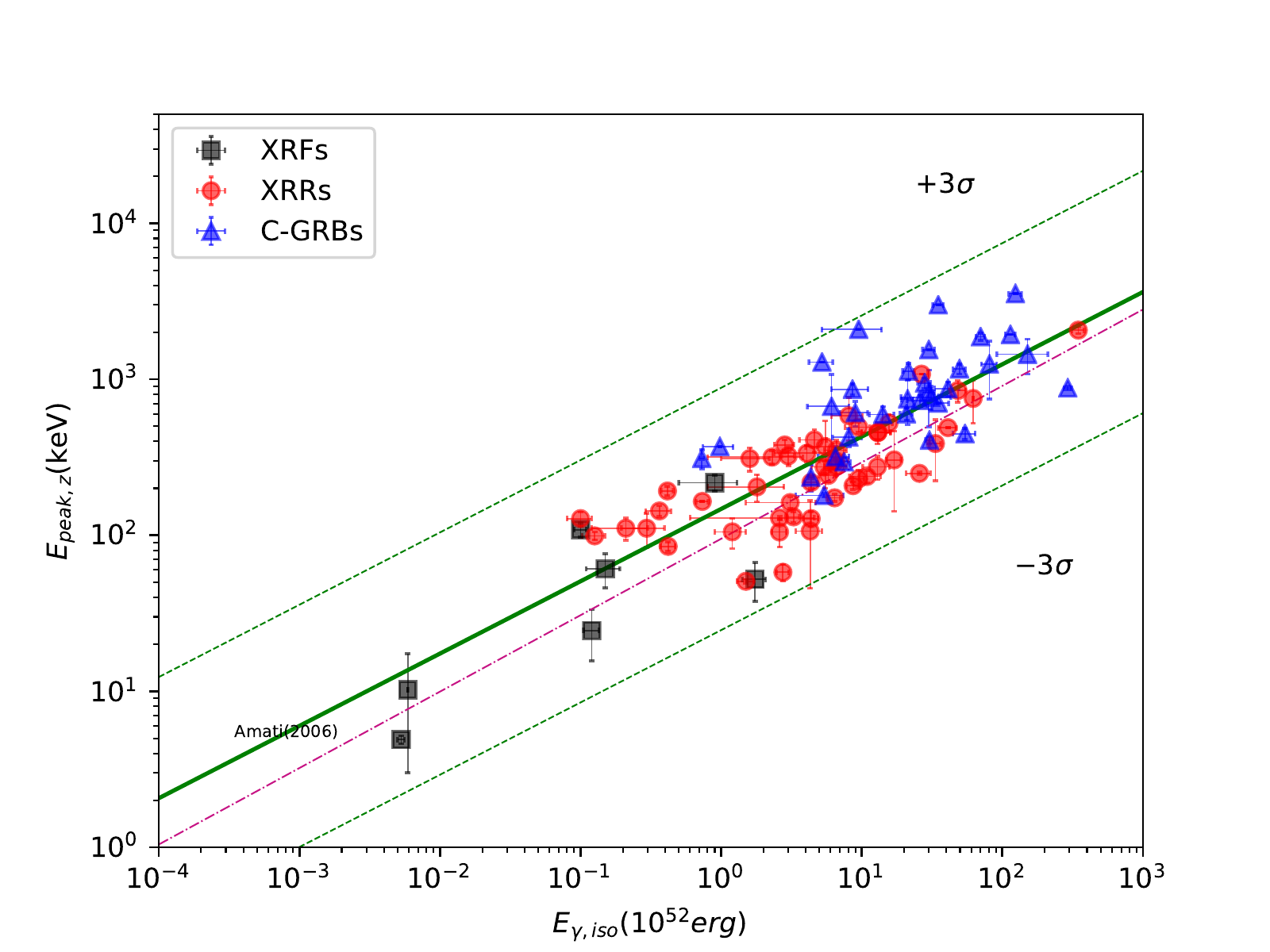}
\caption{Correlation between GRB peak energy $E_{peak,z}$ in the rest frame and GRB isotropic-equivalent energy $E_{\gamma,iso}$. The green solid line is the best fit with the function of
$\log(E_{peak,z})$(keV)=$(2.17\pm 0.04)+(0.46\pm 0.03)\log[E_{\gamma,iso}/(10^{52}erg)]$, and the extrinsic scatter is $\sigma=0.26\pm 0.02$.
and the green dashed lines are marked for the 3$\sigma$ regions. The pink dashed-dotted line is the best fit with the function of $E_{peak,z}$=95keV$(E_{\gamma,iso}/10^{52}erg)^{0.49}$ reported by Amati (2006). XRFs, XRRs, and C-GRBs are marked as black squares, red dots, and blue triangles, respectively.\label{fig10}}
\end{figure}

\begin{figure}
\plotone{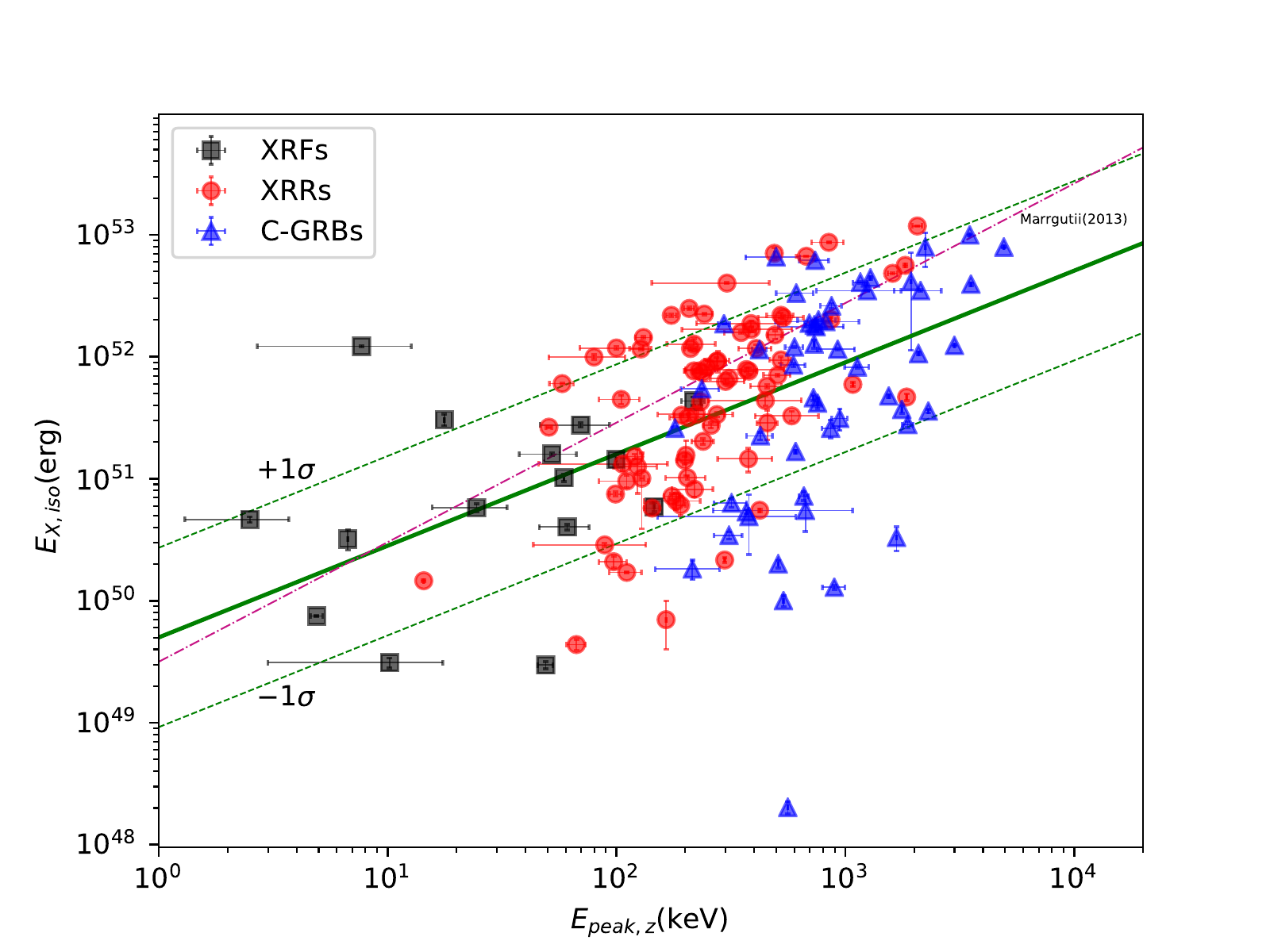}
\caption{Correlation between GRB X-ray energy $E_{X,iso}$ and GRB peak energy $E_{peak,z}$. The green solid line is the best fit with the function of
$log(E_{X,iso})=(49.69\pm 0.27)+(0.75\pm 0.11)log(E_{peak,z})$, and the extrinsic scatter is $\sigma=0.73\pm 0.04$.
The green dashed lines are marked for the 1$\sigma$ regions. The pink dashed line is the best fit using the function of $log(E_{X,iso})=(49.50\pm 0.15)+(0.98\pm 0.02)log(E_{peak,z})$ reported by Margutti et al. (2013). XRFs, XRRs, and C-GRBs are marked as black squares, red dots, and blue triangles, respectively.\label{fig11}}
\end{figure}

\begin{figure}
\plotone{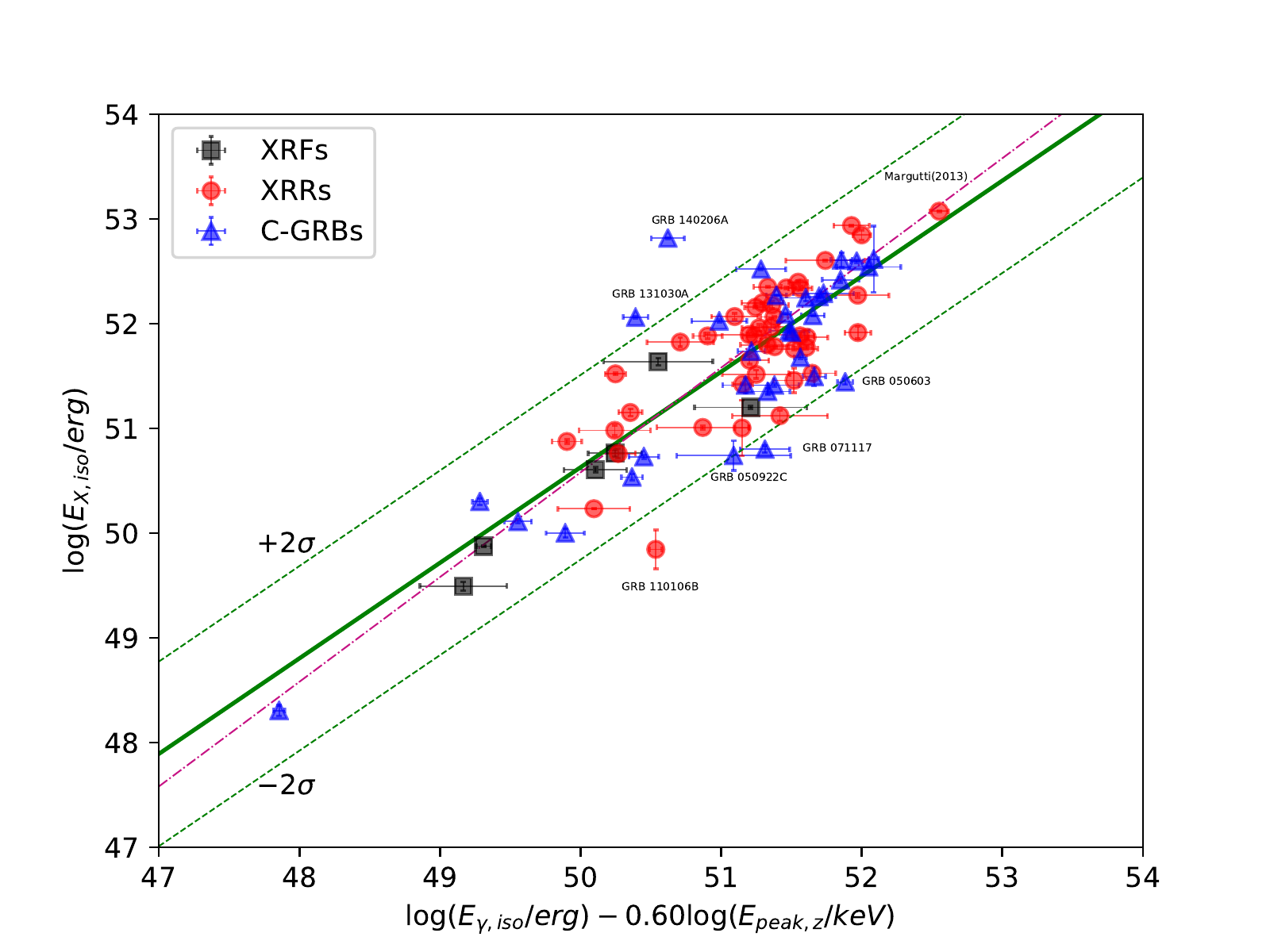}
\caption{Correlation between $E_{X,iso}$ and $E_{\gamma,iso}-(E_{peak,z})$. The green solid line is the best fit with the function of
$log(E_{X,iso})=(4.78\pm 2.79)+(0.92\pm 0.06)(logE_{\gamma,iso}-0.6log(E_{peak})$ with the extrinsic scatter of $\sigma=0.44\pm 0.04$.
The green dashed lines are marked for the 2$\sigma$ regions. The pink dashed-dotted line is the best fit using the function of $log(E_{X,iso})=(0.58\pm 0.25)+(1.00\pm 0.06)(log(E_{\gamma,iso}))-(0.60\pm 0.10)log(E_{peak})$ reported by Margutti et al.(2013).
XRFs, XRRs, and C-GRBs are marked as black squares, red dots, and blue triangles, respectively. All the GRBs labeled at outliers in the figure are L-GRBs. \label{fig12}}
\end{figure}

\begin{figure}
\plotone{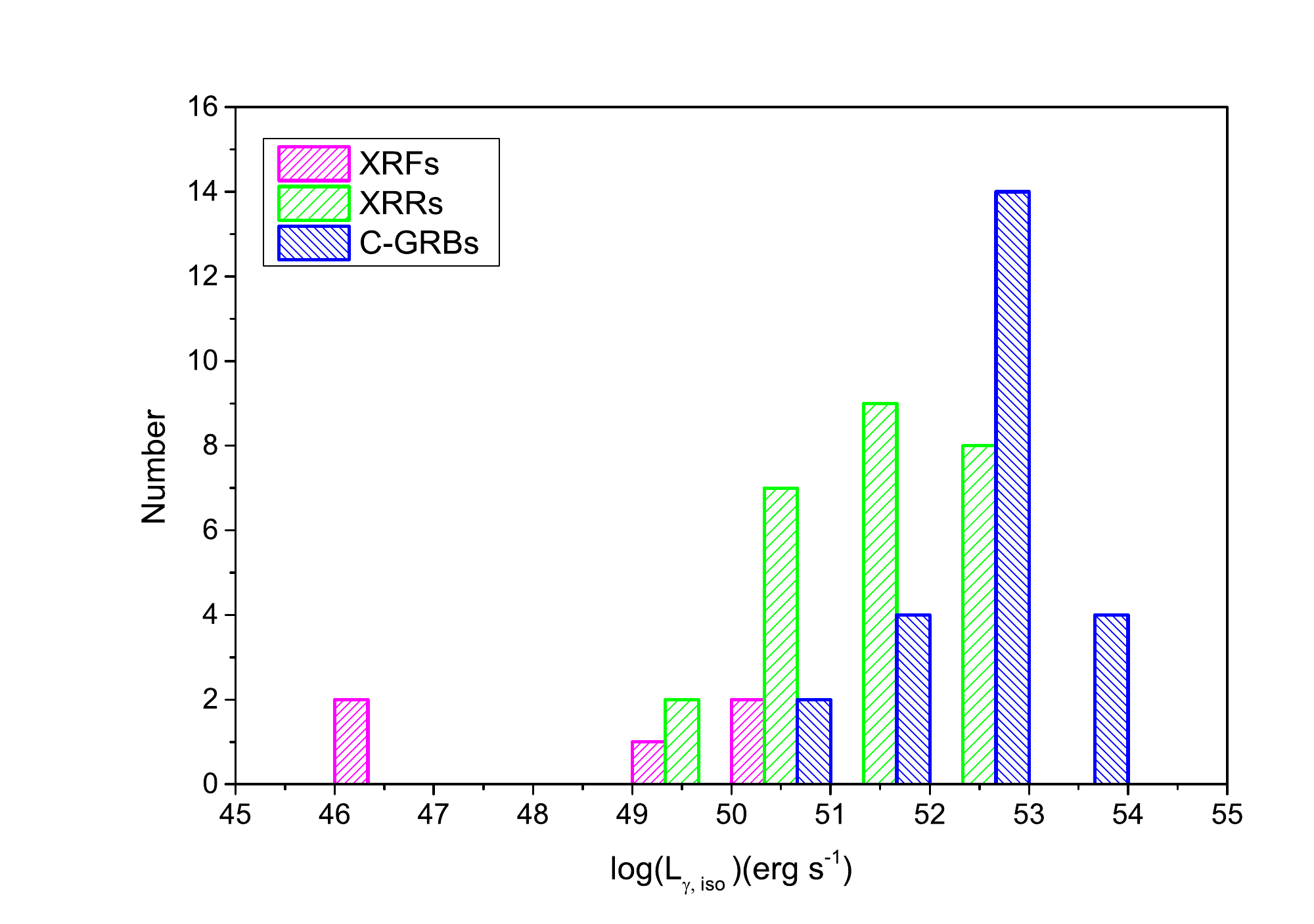}
\caption{GRB isotropic luminosity $L_{\gamma,iso}$ distribution for XRF, XRR, and C-GRB classes. XRFs, XRRs, and C-GRBs are marked as pink, green, and blue, respectively.\label{fig13}}
\end{figure}

\begin{figure}
\plotone{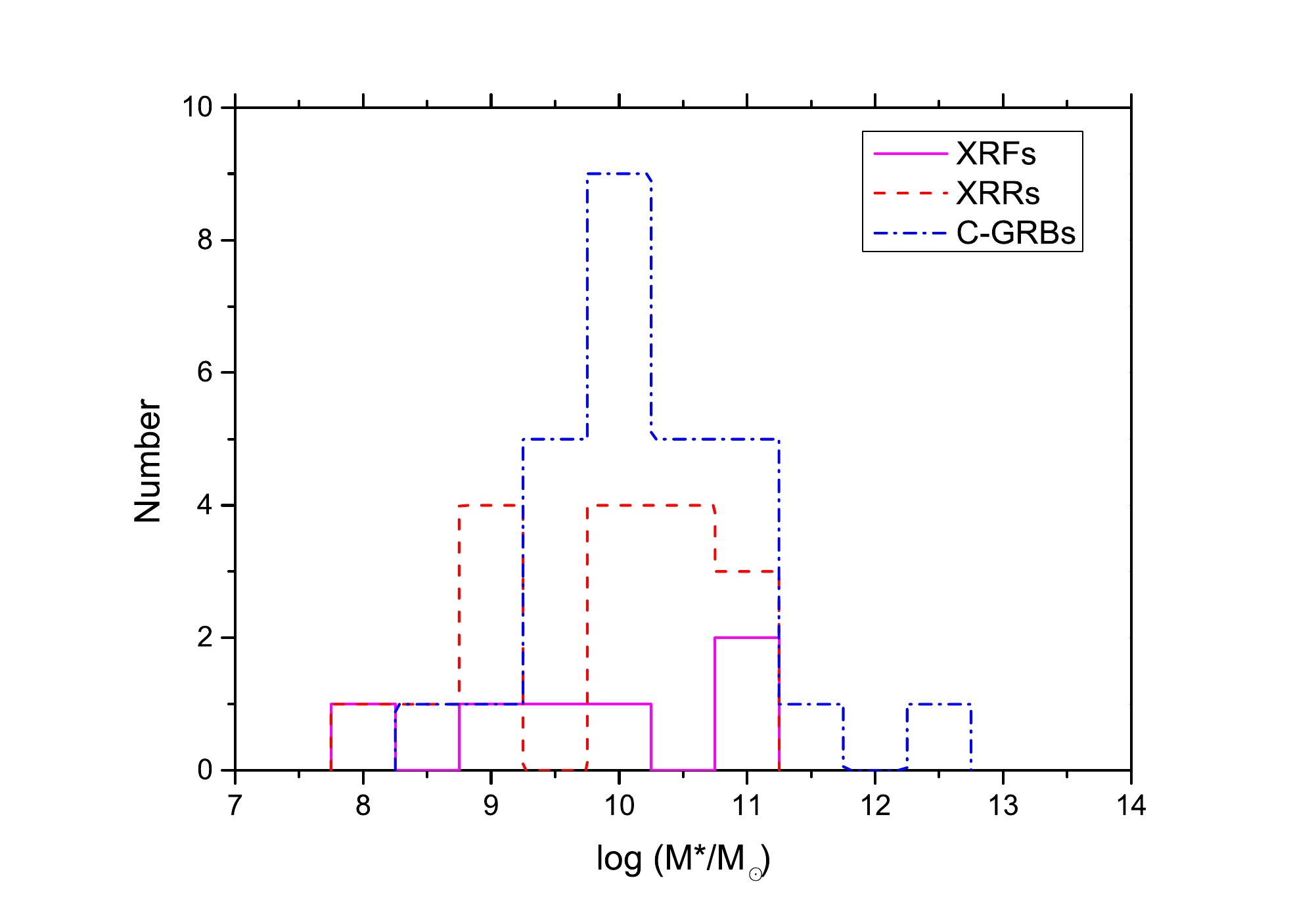}
\caption{Distribution of stellar mass ($M^{\ast}$) in GRB host galaxy for XRFs, XRRs, and C-GRBs. XRFs, XRRs, and C-GRBs are marked as the pink solid line, the red dashed line, and the blue dashed-dotted line, respectively. \label{fig14}}
\end{figure}

\begin{figure}
\plotone{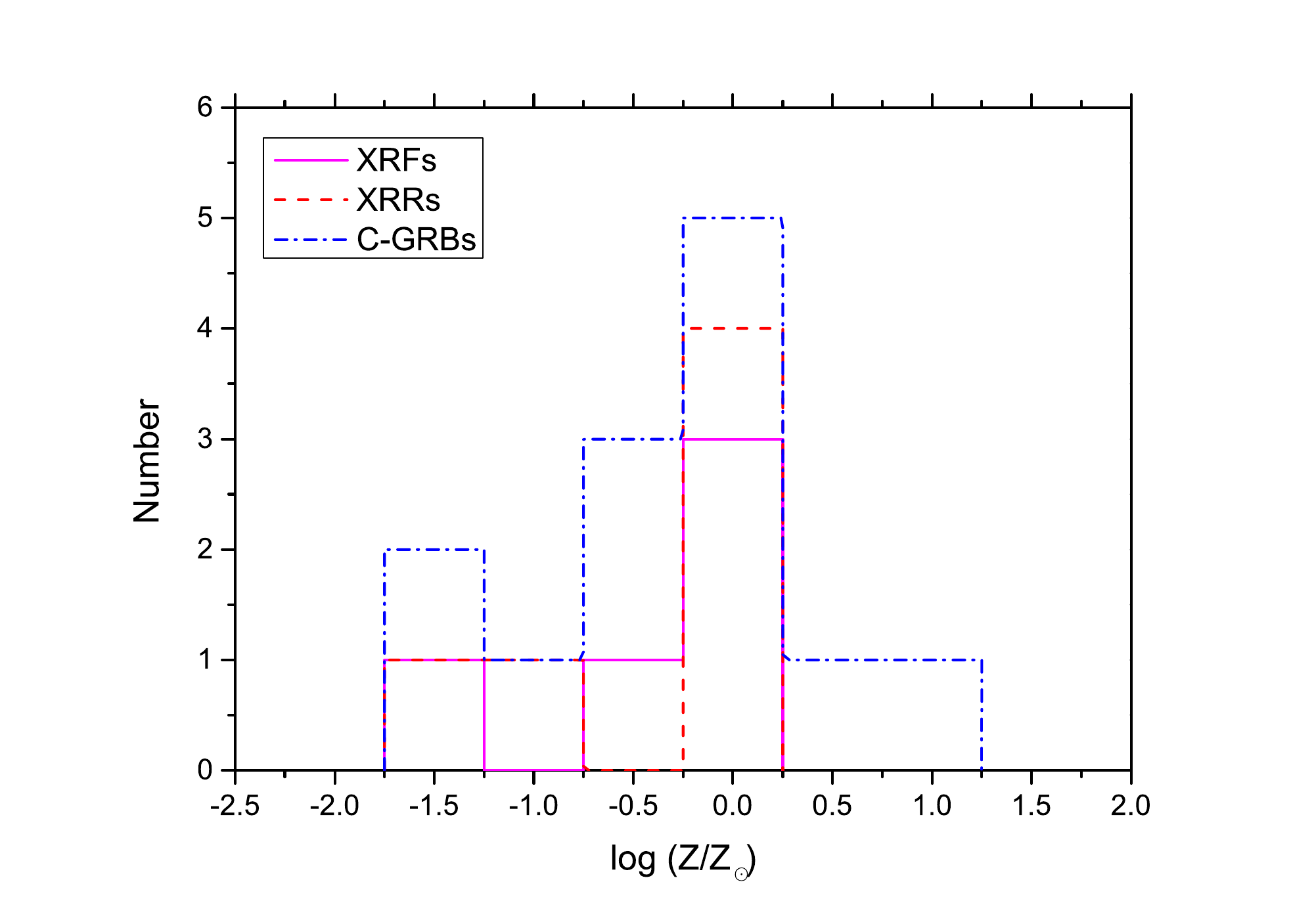}
\caption{Distribution of metallicity ($Z$) in GRB host galaxy for the XRFs, XRRs, and C-GRBs. XRFs, XRRs, and C-GRBs are marked as the pink solid line, the red dashed line, and the blue dashed-dotted line, respectively.\label{fig15}}
\end{figure}

\begin{figure}
\plotone{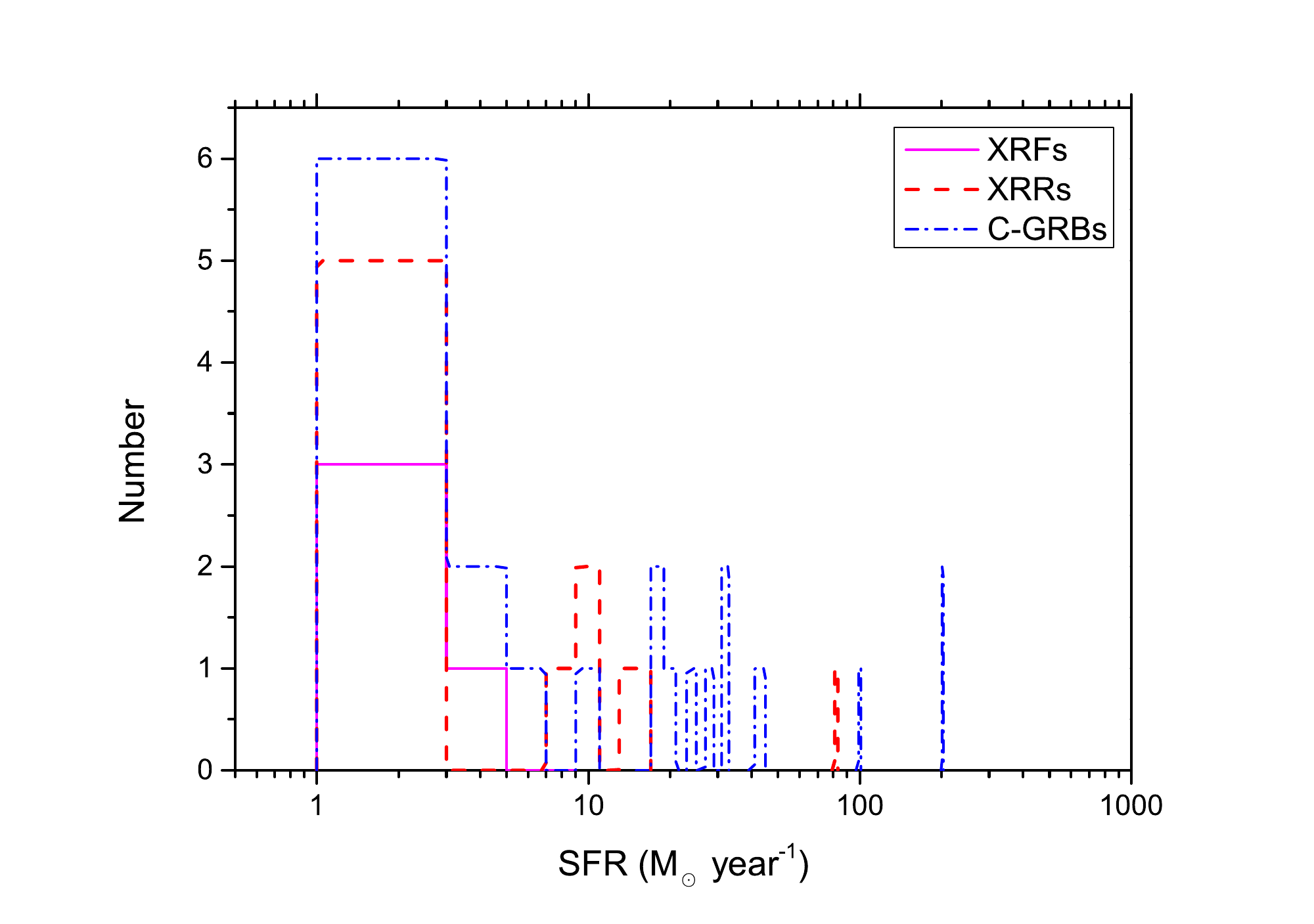}
\caption{Distribution of star formation rate (SFR) in GRB host galaxy for XRFs, XRRs, and C-GRBs. XRFs, XRRs, and C-GRBs are marked as the pink solid line, the red dashed line, and the blue dashed-dotted line, respectively.\label{fig16}}
\end{figure}

\clearpage
\begin{deluxetable}{lcc ccc ccc}
\tabletypesize{\scriptsize}
\tablecaption{Statistical Results of XRF, XRR, and C-GRB in the {\it Swift}-BAT3 GRB catalog\label{tab1}}
\tablewidth{0pt}
\tablehead{
\colhead{Class} &
Total Number [fraction] & L-GRB Number [fraction] & S-GRB Number [fraction] & Unknown$^b$ [fraction]
}
\startdata
XRF & 81 [$(7.3\pm 0.8)$\%] & 70 [$(86.4\pm 10.3)$\%] & 3 [$(3.7\pm 2.1)$\%] & 8 [$(9.9\pm 3.5)$\%] \\
XRR & 540 [$(48.9\pm 2.1)$\%] & 509 [$(94.3\pm 4.2)$\%] & 27 [$(5.0\pm 1.0)$\%] & 4 [$(0.7\pm 0.4)$\%] \\
C-GRB & 394 [$(35.7\pm 1.8)$\%] & 328 [$(83.2\pm 4.6)$\%] & 65 [$(16.5\pm 2.0)$\%] & 1 [$(0.3\pm 0.3)$\%] \\
Unknown$^a$ & 89 [$(8.1\pm 0.9)$\%] & - & - & -  \\
\enddata
\tablecomments{``a" indicates that these sources have no {\it Swift}-XRT observations, and we do not clarify them as XRFs, XRRs, or C-GRBs; ``b" indicates that these sources have no $T_{90}$ numbers, and we cannot judge whether they belong to long-duration GRB or short-duration GRB subclass.}
\end{deluxetable}

\clearpage
\begin{deluxetable}{lcc ccc ccc}
\tablecaption{XRF and XRR samples\label{tab2}}
\tablewidth{0pt}
\tabletypesize{\scriptsize}
\tablehead{
\colhead{GRB Name} & \colhead{z} & \colhead{$T_{90}$} & \colhead{$\alpha_{PL}$} & \colhead{$\alpha_{CPL}$} &\colhead{$S$(15-150 keV)} & \colhead{SR} & \colhead{$E^{obs}_{peak}$} & \colhead{Ref}\\
\colhead{} & \colhead{ } & \colhead{(s)} & \colhead{ } & \colhead{ } &\colhead{$(10^{-8}~\rm{erg~cm^{-2}})$} & \colhead{ } & \colhead{(keV)} & \colhead{ }
}
\startdata
XRF 050215B  &  2.52  &  11.04 &  -2.25$\pm$0.12 &  -1.95 & 22.5 &  1.30$\pm$0.23   &  5.6 &  Lie16 \\
XRF 050406  &  2.7  &  5.78  & -2.40$\pm$0.17 & $0.42^{+1.50}_{-0.89}$ & 6.7$\pm$0.7 &  3.15$\pm$1.22   &  $29.4^{+3.1}_{-3.0}$ &  Lie16 \\
XRF 050416A  &  0.6528  &  6.67 & -3.17$\pm$0.14 & -$0.82^{+1.08}_{-0.77}$ & 37.4$\pm$2.4  &  5.83$\pm$2.39   & $14.8^{+3.6}_{-7.0}$  &  Lie16 \\
XRF 050714B  &  2.438  &  49.36 & -2.40$\pm$0.17 & -$0.01^{+1.04}_{-0.72}$ & 55.0$\pm$4.9  &  2.42$\pm$0.68   &  $30.1^{+3.1}_{-5.2}$   &  Lie16 \\
XRF 050815  & -  &  2.99  & -1.82$\pm$0.12 & $0.40^{+0.86}_{-0.67}$ & 8.2$\pm$0.8  &  1.32$\pm$0.26   & $43.6^{+5.8}_{-3.5}$  &  Lie16 \\
XRF 050819  &  2.5043 &  37.72 & -2.64$\pm$0.14 & -$0.49^{+1.04}_{-0.65}$  & 30.5$\pm$2.4  &  3.08$\pm$0.99   & $23.3^{+3.0}_{-10.7}$   &  Lie16 \\
XRF 050822  &  1.434  &  104.29 & -2.33$\pm$0.07  & -$1.81^{+0.36}_{-0.10}$   & 234.8$\pm$11.5 &  1.43$\pm$0.13  & $13.9^{+7.4}_{-6.5}$  &  Lie16 \\
XRF 050824  &  0.8278  &  25.01 & -2.79$\pm$0.18  &  -1.00 & 25.6$\pm$2.5 &  1.74$\pm$0.79   &  80.0 &  Lie16 \\
\enddata
\tablecomments{$\alpha_{PL}$ is the power-law photon index in the simple power-law model, and the data are derived from Lien et al. (2016). $\alpha_{CPL}$ is the power-law photon index in the cutoff power-law model, and the data are derived from Lien et al. (2016). $S$(15-150 keV) is the fluence of (15-150) keV band, and the data are derived from Lien et al. (2016). $SR$ is the fluence ratio of $S$(25-50 keV)/$S$(50-100 keV). Ref is the reference in which we take $E^{obs}_{peak}$ values: Ama08 refers to Amati, et al. (2008), Ama09 refers to Amati, et al. (2009), Can07 refers to Cano, et al. (2017), Gru13 refers to Grupe et al. (2013), Lia15 refers to Liang et al. (2015), Lie16 refers to Lien et al. (2016), and Zan16 refers to Zaninoni et al. (2016).}
\end{deluxetable}

\clearpage
\begin{deluxetable}{lcc ccc ccc c}
\tabletypesize{\scriptsize}
\tablecaption{Statistics of the X-Ray Lightcurve Type for XRFs, XRRs, and C-GRBs \label{tab3}}
\tablewidth{0pt}
\tablehead{
\colhead{Class} & \colhead
{Total Number} & Type 0 [fraction] & Type  \uppercase\expandafter{\romannumeral1} [fraction] &
Type \uppercase\expandafter{\romannumeral2} [fraction] & Type \uppercase\expandafter{\romannumeral3} [fraction]
}
\startdata
XRF & 34  & 5 [$(14.7\pm 6.6)$\%] & 15 [$(44.1\pm 11.4)$\%] & 11 [$(32.4\pm 9.8)$\%] & 3 [$(8.8\pm 5.1)$\%] \\
XRR & 201 & 35 [$(17.4\pm 2.9)$\%] & 62 [$(30.8\pm 3.9)$\%] & 91 [$(45.3\pm 4.7)$\%] & 13 [$(6.5\pm 1.8)$\%] \\
C-GRB & 135 & 34 [$(25.2\pm 4.3)$\%] & 60 [$(44.4\pm 5.7)$\%] & 37 [$(27.4\pm 4.5)$\%] & 4 [$(3.0\pm 1.5)$\%] \\
\enddata
\end{deluxetable}

\begin{deluxetable}{llc ccc ccc c}
\tabletypesize{\scriptsize}
\tablecaption{Statistical Results of the Association between XRF/XRR and Supernova\label{tab4}}
\tablewidth{0pt}
\tablehead{
 \colhead{Source} & \colhead{$z$} &  \colhead{$E^{obs}_{peak}$(keV)} & \colhead
{Supernova} & \colhead{Grade} & \colhead{Type} & \colhead{$E_{K}$} & \colhead{$log(M^{\ast}/M_{\odot})$} &
\colhead{$log(Z/Z_{\odot})$} & SFR
}
\startdata
XRF 050416A & 0.6528  & $14.8^{+3.6}_{-7.0}$ & \nodata & D & \nodata & \nodata  & 9.19 & \nodata & 2.32 \\
XRR 050525A & 0.606  & $80.4^{+1.7}_{-1.5}$  & 2005nc  & B &  \nodata & $1.89^{+1.07}_{-0.75}$ & \nodata & \nodata & \nodata \\
XRF 050824 & 0.8278 & 80.0 & \nodata & E & \nodata & $0.57^{+0.93}_{-0.37}$ & \nodata  & -0.3 & \nodata \\
XRF 060218 & 0.03342 & 4.7$\pm$0.3 & 2006aj & A & Ic & 1.02$\pm$0.23 & 7.78 & -0.53 & 0.05 \\
XRR 060729 & 0.5428 & 201.2$\pm$68.9 & \nodata & D & \nodata & $2.44^{+1.43}_{-0.99}$ & \nodata & \nodata & \nodata \\
XRR 060904B & 0.7029 & 84.1$\pm$13.3 & \nodata & C & \nodata & $0.99^{+0.51}_{-0.37}$ & \nodata  & \nodata  & \nodata  \\
XRF 070419A & 0.9705 & $26.5^{+3.6}_{-11.2}$ & \nodata & D & \nodata & \nodata & \nodata & \nodata & \nodata\\
GRB 080319B & 0.9382 & 650.6$\pm$33.5 &  \nodata & C & \nodata & $2.27^{+1.91}_{-1.19}$ & \nodata & \nodata & \nodata\\
\enddata
\tablecomments{Grades are classified by Hjorth \& Bloom (2012): (A) Strong spectroscopic evidence; (B) A clear light curve bump as well as some spectroscopic evidence resembling a GRB-SN; (C) A clear bump consistent with other GRB-SNe at the spectroscopic redshift of the GRB; (D) A bump, but the inferred SN properties are not fully consistent with other GRB-SNe or the bump was not well sampled or there is no spectroscopic redshift of the GRB; (E) A bump, either of low significance or inconsistent with other GRB-SNe. Type refers to the supernova explosion type. $E_{K}$ is the supernova ejecta kinetic energy, in the unit of $10^{52}$ erg. $M^{\ast}$ is the stellar mass of the GRB host galaxy. $Z$ is the metallicity of the GRB host galaxy. SFR is the star formation rate of the GRB host galaxy, in the unit of $M_{\odot}~\rm{yr^{-1}}$.}
\end{deluxetable}

\clearpage
\begin{deluxetable}{llc ccc}
\tabletypesize{\scriptsize}
\tablecaption{Host Galaxy Properties of XRFs and XRRs \label{tab5}}
\tablewidth{0pt}
\tablehead{
\colhead{Source} & \colhead{$z$} & \colhead{$E^{obs}_{peak}$(keV)} & SFR & \colhead{$log(M^{\ast}/M_{\odot})$} & \colhead{$log(Z/Z_{\odot})$}
}
\startdata
XRF 050416A & 0.6528  & $14.8^{+3.6}_{-7.0}$ & 2.32& 9.19 & - \\
XRF 050824 & 0.8278 & 80.0 & - & - & -0.3  \\
XRF 060218 & 0.03342 & 4.7$\pm$0.3 & 0.05 & 7.78 & -0.53 \\
XRF 061210 & 0.4095 & - & - & 9.52 & - \\
XRF 071031 & 2.692 & $2.1^{+187.8}_{-107.5}$  & - & - & -1.85 \\
XRF 071227 & 0.381 & 35.5$\pm$5.3 & 0.6 & 10.65 & -0.2 \\
XRF 090205 & 4.6497 & $38.4^{+5.0}_{-3.8}$ & - & 10.83 & -  \\
XRF 100316D & 0.0591 & $9.6^{+9.3}_{-4.3}$ & 0.14 & 8.93 & -0.39  \\
\enddata
\tablecomments{SFR is star formation rate of GRB host galaxy, in the unit of $M_{\odot}~yr^{-1}$. $M^{\ast}$ is stellar mass of GRB host galaxy. $Z$ is metallicity of GRB host galaxy. All data are taken from GRB Host Studies (GHostS) database.}
\end{deluxetable}



\end{document}